\documentclass[a4paper,10pt]{article}
\pdfoutput=1 

\usepackage{jheppub} 

\usepackage[T1]{fontenc} 
\usepackage{lmodern} 

\usepackage{xcolor}
\usepackage{mathrsfs,wasysym}
\usepackage{booktabs}
\usepackage{slashed}
\usepackage{cancel}

\newcommand{\kt}{k_{\rm t}}
\newcommand{\ktn}[1]{k_{{\rm t} #1}}

\newcommand{\MSbar}{\overline{\text{MS}}}
\newcommand{\sss}{\scriptscriptstyle}

\newcommand{\rhot} {\rho_t}

\newcommand{\as}{\alpha_s}

\newcommand{\Ord}{\mathcal{O}}
\newcommand{\Lum}{\mathscr{L}}

\newcommand{\mh}{m_{H}}
\newcommand{\mt}{m_{t}}

\newcommand{\muf}{\mu_{\sss\rm F}}
\newcommand{\mur}{\mu_{\sss\rm R}}

\newcommand{\Li}{\mathrm{Li}}

\newcommand{\abs}[1]{\left| #1 \right|}

\let\originalleft\left
\let\originalright\right
\renewcommand{\left}{\mathopen{}\mathclose\bgroup\originalleft}
\renewcommand{\right}{\aftergroup\egroup\originalright}

\def\beq{\begin{equation}}  
\def\eeq{\end{equation}}
\def\({\left(}
\def\){\right)}
\def\[{\left[}
\def\]{\right]}

\let\oldsubsection\subsection
\renewcommand\subsection[2][\subsectiontoc]{%
  \def\subsectiontoc{#2}%
  \oldsubsection[#1]{\boldmath #2}%
}

\let\oldsubsubsection\subsubsection
\renewcommand\subsubsection[2][\subsubsectiontoc]{%
  \def\subsubsectiontoc{#2}%
  \oldsubsubsection[#1]{\boldmath #2}%
}

\allowdisplaybreaks

\newcommand{\new}[1]{\color{black}#1\color{black}}

\arxivnumber{1805.08785}

\title{\boldmath Small-$x$ phenomenology at the LHC and beyond: HELL~3.0 and the case of the Higgs cross section}

\author[]{Marco Bonvini}
\affiliation[]{INFN, Sezione di Roma 1,\\ Piazzale Aldo Moro~5, 00185 Roma, Italy}

\preprint{}

\emailAdd{marco.bonvini@roma1.infn.it}

\abstract{%
Small-$x$ resummation has been proven recently to be a crucial ingredient for describing small-$x$ HERA data,
and the inclusion of small-$x$ resummation in parton distribution function (PDF) determination
has a sizeable effect on the PDFs even at the electroweak scale.
In this work we explore the implications of small-$x$ resummation at the Large Hadron Collider (LHC)
and at a Future Circular Collider (FCC).
We construct the theoretical machinery for resumming physical inclusive observables at hadron colliders,
and describe its implementation in the public code \texttt{HELL~3.0}.
We focus on Higgs production in gluon fusion as a prototypical example,
both because it is sensitive to small-$x$ gluons and because of its importance for the LHC physics programme.
We find that adding small-$x$ resummation to the N$^3$LO Higgs production cross section can lead to an increase of up to $10\%$
at FCC, while the effect is smaller ($+1\%$) at LHC but still important to achieve a high level of precision.
}

\begin{document}

\maketitle

\section{Introduction}

With the discovery of the Higgs boson, the Standard Model (SM) has been established as a successful theory of particle physics.
While the SM cannot be the definitive theory, direct evidence of physics beyond the SM has not (yet) been observed at the LHC.
The search for new phenomena beyond the SM at hadron colliders may be pursued by testing the SM to high precision,
which is becoming possible thanks to the huge amount and excellent quality
of the data collected by the LHC.
To keep up, theoretical predictions must reach and possibly surpass the precision of the measurements.
On the one hand, this requires refined theoretical predictions for the partonic cross sections for the processes of interest,
which may be obtained by higher order computations, e.g.\ next-to-next-to-leading order (NNLO)
or even next-to-next-to-next-to-leading order (N$^3$LO) in some cases,
and by the all-order resummation of important classes of logarithmic contributions.
On the other hand, accurate and precise theoretical predictions for LHC processes require
high-quality parton distribution functions (PDFs).

Recently, an important step forward towards improved determination of PDFs
has been achieved in Refs.~\cite{Ball:2017otu,Abdolmaleki:2018jln}, where the resummation of small-$x$
(high-energy) logarithms at next-to-leading logarithmic (NLL) accuracy
as implemented in the public code \texttt{HELL}~\cite{Bonvini:2016wki,Bonvini:2017ogt}
has been included in PDF evolution and in the theoretical predictions of DIS observables.
Small-$x$ resummation has the important role of stabilizing the behaviour of DGLAP splitting functions 
at small $x$, which otherwise is compromised by powers of $\log\frac1x$.
In particular, the first manifest instability appears at NNLO, and thus 
PDFs determined with NNLO theory are rather different to those determined
with NNLO theory improved by NLL small-$x$ resummation.
This difference, determined at low $Q^2$ where the small-$x$ HERA data lie, persists and
is actually enlarged by DGLAP evolution at larger scales. As a result, resummed PDFs
at the electroweak scale are very different from the NNLO ones at small~$x$.

This raises an important question: how does this large effect impact LHC precision phenomenology?
To properly answer, we need to compare fixed-order prediction with fixed-order (NNLO) PDFs
to resummed predictions with resummed (NNLO+NLL) PDFs.
While NNLO+NLL PDFs are now available, resummed predictions for LHC observables did not exist,
or at least not in a format which makes them immediately usable for phenomenology.
It is the goal of this paper to provide the theoretical setup to perform this resummation
for inclusive observables with the public code \texttt{HELL}.
The resummation of differential observables with \texttt{HELL} is left to future work.

As a first example of application of this setup, we will consider Higgs production in gluon fusion.
Being initiated by two initial-state gluons, this process is very sensitive to the gluon PDF.
Moreover, it is known that the inclusive Higgs cross section is dominated by contributions
close to partonic threshold, which in turn implies that the gluon PDF contributes mostly at small $x$.
In addition, the inclusive Higgs cross section in gluon fusion is known to N$^3$LO~\cite
{Anastasiou:2015ema,Anzai:2015wma,Anastasiou:2016cez,Mistlberger:2018etf},
so we will provide all the ingredients to properly match small-$x$ resummation
of a physical process to N$^3$LO for the first time.
We then investigate the phenomenological implications of small-$x$ resummation in Higgs production
at the LHC, and to enlarge the sensitivity to the PDFs at small $x$ also at higher-energy colliders,
namely High-Energy LHC (HE-LHC) and a Future Circular hadron-hadron Collider (FCC-hh).

The structure of the paper is the following.
In Sect.~\ref{sec:SXhh} we derive the formalism for small-$x$ resummation of inclusive cross sections
with two hadrons in the initial state. We discuss its implementation in the \texttt{HELL} code,
and compare it to the original formulation~\cite{Ball:2007ra} in the Altarelli-Ball-Forte (ABF) formalism~\cite
{Ball:1995vc,Ball:1997vf,Altarelli:2001ji,Altarelli:2003hk,Altarelli:2005ni,Altarelli:2008aj}.
We provide all the ingredients for matching small-$x$ resummation in the partonic coefficient functions to N$^3$LO.
In Sect.~\ref{sec:ggH} we move to Higgs production, and present first how the fixed-order
cross section can be constructed to treat correctly the small-$x$ behaviour at NNLO and N$^3$LO,
and then the effect of adding small-$x$ resummation both at parton level and at the level of the physical cross section.
We then draw our conclusions in Sect.~\ref{sec:conclusions}, and collect technical details in App.~\ref{sec:offshell}.
This work represents a follow up of Refs.~\cite{Bonvini:2016wki,Bonvini:2017ogt,Bonvini:2018xvt} and \cite{Ball:2017otu},
and provides the foundations of Ref.~\cite{Bonvini:2018ixe}.

\section{Hadron-hadron collider processes at high-energy}
\label{sec:SXhh}

The resummation of small-$x$ logarithms in physical processes requires both using PDFs which include small-$x$ resummation in their determination and evolution,
and resumming to all orders the $\log\frac1x$ contributions in the partonic coefficient functions.
The latter resummation, which is the subject of this section, is based on the so-called $\kt$ factorization theorem,
where the non-perturbative proton dynamics is factorized in parton distribution functions which depend on both
the longitudinal momentum fraction $x$ of the parton and its transverse momentum $\kt$~\cite{Catani:1990xk,Catani:1990eg,Collins:1991ty,Catani:1993ww,Catani:1993rn,Catani:1994sq}.
Relating this $\kt$-dependent PDFs to the usual collinear PDFs it is possible to resum the leading non-vanishing tower of small-$x$ logarithms to all orders
in the collinearly factorized partonic coefficient functions.

Another important ingredient for a stable small-$x$ resummation is the inclusion to all orders
of a class of subleading contributions originating from the running
of the strong coupling $\as$~\cite{Ball:2007ra,Altarelli:2008aj}.
In Ref.~\cite{Bonvini:2016wki} the approach of Refs.~\cite{Ball:2007ra,Altarelli:2008aj} has been rederived and reformulated in a simpler and more general way,
and proven to be identical to the original formulation under specific assumptions.
The new formulation of Ref.~\cite{Bonvini:2016wki} has been implemented
in the computer code \href{http://www.ge.infn.it/~bonvini/hell}{\texttt{HELL}}~\cite{Bonvini:2016wki,Bonvini:2017ogt},
and it is very convenient from the analytical and numerical points of view,
making the resummation of new processes and their inclusion in \texttt{HELL} rather straightforward.
In Ref.~\cite{Bonvini:2016wki}, and subsequently in Ref.~\cite{Bonvini:2017ogt}, this new formalism has been presented and used
only for processes with a single hadron in the initial state, and specifically the deep inelastic scattering (DIS) process.
In this section we extend the formulation to processes with two hadrons in the initial state,
relevant for hadron-hadron colliders such as the LHC.
This extension was already presented in the orignal formulation in Ref.~\cite{Ball:2007ra,Marzani:2008uh};
in this section we will also show that our formulation, which is more general,
reduces to the original one under the same assumptions considered for the single-hadron case.

\subsection{Resummation formalism with two incoming gluon legs}

\begin{figure}[t]
  \centering
  \includegraphics[width=0.33\textwidth]{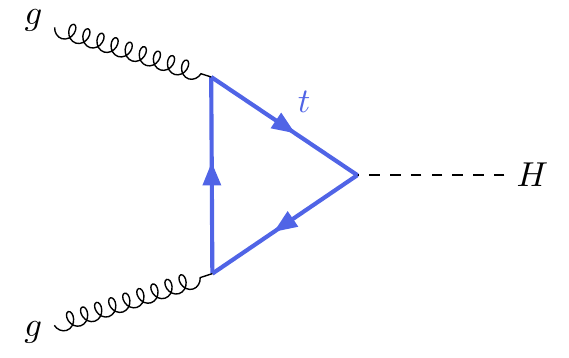}
  \caption{Leading order diagram for Higgs production in gluon fusion at hadron-hadron colliders. The quark running in the loop is predominantly a top.}
  \label{fig:ggHprocLO}
\end{figure}

We consider a hadron-collider process which is gluon-gluon initiated.
The typical and cleanest example, which we will consider in greater detail later in Sect.~\ref{sec:ggH}, is Higgs production in gluon fusion,
whose leading order diagram is depicted in Fig.~\ref{fig:ggHprocLO}.
Other examples for which the results of this section will be relevant are, e.g., top-pair production and jet production.

We will assume that there are no collinear singularities to be subtracted at resummed level.
Namely, the lowest order diagram with two gluons in the initial state must be finite without any collinear subtraction.
Indeed, in order for a collinear singularity to be present, at least one of the gluons must split into a pair of quarks,
one of which participates to the hard interaction.
In other words, it must be possible to cut a quark line such that the diagram factorizes into a gluon splitting to quarks
and a gluon-quark initiated subgraph.
Therefore, in presence of collinear singularities in a $gg$ initiated diagram, there must exist a lower order diagram which is $gq$ initiated.
But if this is the case, the resummation of the $gg$ initiated process is subleading logarithmic with respect to the resummation
of the $gq$ initiated process, due to the extra power of $\as$ and no logarithm in the $g\to q\bar q$ splitting.
Thus, at the leading non-vanishing logarithmic accuracy, contribution with two initial state gluons which require collinear subtractions do not contribute.
This is the case for instance of Drell-Yan production, where indeed at lowest logarithmic order only the $gq$ (and $qq$) channels contribute~\cite{Marzani:2008uh}.
Of course, it is well possible that such $gq$ channel contains itself a collinear singularity (as it happens in the Drell-Yan case).
However, this process has a single gluon in the initial state, and the treatment is identical to the DIS case already discussed in Ref.~\cite{Bonvini:2016wki}.

Let us then focus on the cross section $\sigma$ of a gluon-gluon initiated (at lowest order) process
without collinear singularities, such as Higgs production, in hadron-hadron collision.
In order to simplify the treatment, we take the Mellin transform of the cross section as
\beq
\sigma(N,Q^2) = \int_0^1 d\tau\, \tau^N\, \sigma(\tau,Q^2),
\eeq
where $\tau=Q^2/s$, with $Q$ the hard scale of the process (e.g., the Higgs mass) and $\sqrt{s}$ the collider center-of-mass energy.
The cross section in collinear factorization can be written in Mellin $N$ space as the sum over partonic channels of simple products,
\beq\label{eq:sigmaColl}
\sigma(N,Q^2) = \sigma_0(N,Q^2)
\sum_{i,j=g,q} C_{ij}\(N,\as,\frac{\muf^2}{Q^2}\)\, f_i(N,\muf^2)\, f_j(N,\muf^2),
\eeq
where $C_{ij}$ are the collinearly factorized coefficient functions and $f_{k}$ the collinear PDFs, which depend on the factorization scale $\muf\sim Q$.
The strong coupling $\as$ is in general evaluated at the renormalization scale $\mur$,
which implies that there are logarithms of $\mur/Q$ in the coefficient function to compensate its dependence;
however, at the leading logarithmic accuracy we will consider, the $\mur$ dependence is subleading, and we therefore omit it to simplify the notation.
The factor $\sigma_0$ is chosen such that the coefficients functions are dimensionless, and normalized to $1$ at LO in the dominant channel
(supposed to be the $gg$ channel in our case).
In the high-energy limit, there is no need to distinguish the individual quarks, as they always contribute in the singlet combination.
Thus, in this section, we will assume that the index $q$ refers to the whole singlet PDF.\footnote
{With this assumption Eq.~\eqref{eq:sigmaColl} is incomplete as it misses non-singlet contributions;
  however, this is irrelevant for the present discussion, which is focussed on the high-energy limit.}

In the high-energy limit, the cross section can be also written according to the $\kt$ factorization theorem,
which gives
\beq\label{eq:sigmakt}
\sigma(N,Q^2) = \sigma_0(N,Q^2)
\int d\ktn1^2\, d\ktn2^2\; {\cal C}\(N, \frac{\ktn1^2}{Q^2}, \frac{\ktn2^2}{Q^2},\as\) \, {\cal F}_g(N,\ktn1^2)\, {\cal F}_g(N,\ktn2^2).
\eeq
Here, ${\cal F}_g$ is the $\kt$-dependent gluon PDF, and ${\cal C}$ is the partonic coefficient function computed with two off-shell incoming gluons,
the off-shellness being $\kt^2$. Obviously, the off-shell coefficient function is symmetric for the exchange of the two virtualities, $\ktn1^2\leftrightarrow\ktn2^2$.
The $\kt$-dependent gluon PDF can be related to the collinear PDFs through the relation~\cite{Bonvini:2016wki, Marzani:2008uh, Catani:1994sq}
\beq\label{eq:Foff}
{\cal F}_g(N,\kt^2) = {\cal U}\(N,\kt^2,\muf^2\) f_g(N,\muf^2) + \frac{C_F}{C_A} \[{\cal U}\(N,\kt^2,\muf^2\) - \delta(\kt^2)\] f_q(N,\muf^2),
\eeq
where ${\cal U}$ is a function which is factorization scheme dependent.\footnote
{We observe that in Ref.~\cite{Bonvini:2016wki} the same equation was written in terms of the ``plus'' eigenvector PDF, see Eq.~(3.3) there.
However, that expression misses a contribution $+\delta(\kt^2)f_-(N,Q^2)$ in terms of the ``minus'' eigenvector PDF, which produces the
$\delta(\kt^2)$ term in Eq.~\eqref{eq:Foff}. The results of Ref.~\cite{Bonvini:2016wki} are unaffected;
the only effect of that deficiency is that $C_-$ appearing in Eq.~(3.26) could be written in terms of the off-shell coefficient function.
However, that contribution is purely NLO, and could thus be extracted from the fixed-order computation.}
In the $Q_0\MSbar$ scheme~\cite{Catani:1993ww,Catani:1994sq,Ciafaloni:2005cg,Marzani:2007gk}
usually considered in the high-energy regime, and adopted also here, it is given by
\beq
{\cal U}_{Q_0\MSbar}\(N,\kt^2,\muf^2\) = \frac{d}{d\kt^2} U(N,\kt^2,\muf^2),
\eeq
with~\cite{Bonvini:2016wki}
\beq\label{eq:Udef}
U(N,\kt^2,\muf^2) = \exp\int_{\muf^2}^{\kt^2} \frac{d\mu^2}{\mu^2} \gamma_+\(N,\as(\mu^2)\),
\eeq
where $\gamma_+$ is the (small-$x$ resummed) eigenvalue of the anomalous dimension singlet matrix which is singular at small $x$.
With respect to Ref.~\cite{Bonvini:2016wki}, we are slightly changing the notation for the evolution function $U$,
to extend it to the case in which $\muf$ is different from the hard scale $Q$.
More details on the actual form of the evolution function $U$ and on the anomalous dimension used in its definition are given later in Sect.~\ref{sec:evolU}.

Plugging Eq.~\eqref{eq:Foff} into Eq.~\eqref{eq:sigmakt} and comparing with Eq.~\eqref{eq:sigmaColl},
we find a relation between the coefficient functions in collinear factorization and the off-shell coefficient functions,
\begin{subequations} \label{eq:CofcalC}
\begin{align}
  C_{gg}\(N,\as,\frac{\muf^2}{Q^2}\)
  &= \int d\xi_1\, d\xi_2\; {\cal C}\(N, \xi_1, \xi_2,\as\) \, \frac{d}{d\xi_1} U(N,Q^2\xi_1,\muf^2) \, \frac{d}{d\xi_2} U(N,Q^2\xi_2,\muf^2), \label{eq:CggofcalC} \\
  C_{qg}\(N,\as,\frac{\muf^2}{Q^2}\)
  &= \frac{C_F}{C_A}\int d\xi_1\, d\xi_2\; {\cal C}\(N, \xi_1, \xi_2,\as\) \, \frac{d}{d\xi_1} U(N,Q^2\xi_1,\muf^2) \, \[\frac{d}{d\xi_2} U(N,Q^2\xi_2,\muf^2) - \delta(\xi_2) \], \\
  C_{qq}\(N,\as,\frac{\muf^2}{Q^2}\)
  &= \(\frac{C_F}{C_A}\)^2\int d\xi_1\, d\xi_2\; {\cal C}\(N, \xi_1, \xi_2,\as\) \, \[\frac{d}{d\xi_1} U(N,Q^2\xi_1,\muf^2)-\delta(\xi_1)\] \nonumber\\
&\hspace{14.8em}\times \[\frac{d}{d\xi_2} U(N,Q^2\xi_2,\muf^2) - \delta(\xi_2) \],
\end{align}
\end{subequations}
where we have introduced the dimensionless variables
\beq
\xi = \frac{\kt^2}{Q^2}.
\eeq
Introducing the ``auxiliary'' coefficient function
\begin{align}\label{eq:Caux}
C_{\rm aux}\(N,\as,\frac{\muf^2}{Q^2}\)
&= -\int d\xi_1\, d\xi_2\; {\cal C}\(N, \xi_1, \xi_2,\as\) \, \frac{d}{d\xi_1} U(N,Q^2\xi_1,\muf^2) \, \delta(\xi_2) \nonumber\\
&= -\int d\xi\; {\cal C}\(N, \xi, 0, \as\) \, \frac{d}{d\xi} U(N,Q^2\xi,\muf^2),
\end{align}
we can rewrite the quark coefficient functions as
\begin{subequations} \label{eq:CofCggaux}
\begin{align}
  C_{qg}\(N,\as,\frac{\muf^2}{Q^2}\)
  &= \frac{C_F}{C_A} \[ C_{gg}\(N,\as,\frac{\muf^2}{Q^2}\) + C_{\rm aux}\(N,\as,\frac{\muf^2}{Q^2}\)\], \\
  C_{qq}\(N,\as,\frac{\muf^2}{Q^2}\)
  &= \(\frac{C_F}{C_A}\)^2 \[ C_{gg}\(N,\as,\frac{\muf^2}{Q^2}\) + 2 C_{\rm aux}\(N,\as,\frac{\muf^2}{Q^2}\) + {\cal C}\(N, 0, 0, \as\)\].
\end{align}
\end{subequations}
These expressions, already derived e.g.\ in Ref.~\cite{Harlander:2009my}, allow us to express both quark coefficient functions
in terms of the gluon one and of the auxiliary function.
Thus, from now on we will focus on the functions $C_{gg}$, Eq.~\eqref{eq:CggofcalC}, and $C_{\rm aux}$, Eq.~\eqref{eq:Caux}.

\subsection{The evolution function}
\label{sec:evolU}

The evolution function $U$, Eq.~\eqref{eq:Udef}, is a key object for small-$x$ resummation in partonic coefficient functions.
Indeed Eqs.~\eqref{eq:CofcalC} encode small-$x$ resummation thanks to the form of $U$,
which contains the leading small-$x$ logarithms to all orders, provided the anomalous dimension in there is itself accurate at least at LL.
We thus now recall here some properties of this function already presented in Refs.~\cite{Bonvini:2016wki,Bonvini:2017ogt},
with particular focus on its $\muf$ dependence that we are now including.

First, we observe that the anomalous dimension in Eq.~\eqref{eq:Udef} is integrated between $\muf$ and $\kt$, and $\kt$
is integrated in Eqs.~\eqref{eq:CofcalC} over all accessible values.
This means that the resummed anomalous dimension would be needed at all possible values of $\as$ between zero and infinity,
which represents a big numerical challenge.
In order to avoid this problem, an approximation of the $\as$ dependence of the anomalous dimension
was proposed in Ref.~\cite{Bonvini:2016wki}, where
\beq\label{eq:gammaasapprox}
\gamma_+(N,\as(\mu^2)) \simeq \frac{\gamma_+(N,\as(\muf^2))}{1+r(N,\as(\muf^2))\log(\mu^2/\muf^2)},\qquad
r(N,\as) = \as^2\beta_0\frac{d}{d\as}\log\[\gamma_+(N,\as)\],
\eeq
with $\beta_0$ the one-loop coefficient of the QCD $\beta$-function.
Under this assumption, the evolution function becomes
\beq
U(N,Q^2\xi,\muf^2) \simeq U_{\rm ABF}\(N,\frac{Q^2}{\muf^2}\xi\),
\eeq
having defined
\beq\label{eq:UABF}
U_{\rm ABF}(N,\zeta) = \Big(1+r(N,\as)\log\zeta\Big)^{\gamma_+(N,\as)/r(N,\as)}.
\eeq
Note that $\as$ in Eq.~\eqref{eq:UABF} is in principle $\as(\muf^2)$;
however, since the scale dependence of $\as$ is subleading with respect to the leading logarithmic accuracy of the resummed coefficient functions,
$\as$ can be computed at any renormalization scale $\mur$ without compensating for this change.
The name ABF in Eq.~\eqref{eq:UABF} comes from the fact that with this approximated evolution function the approach of Refs.~\cite{Altarelli:2008aj,Ball:2007ra}
is recovered, as proven in Ref.~\cite{Bonvini:2016wki} for DIS.
We will show in the next Sect.~\ref{sec:equivalence} that this is the case also for processes with two incoming gluons.

A second observation is related to the region of $\xi$ accessible in the integrals Eqs.~\eqref{eq:CofcalC} and \eqref{eq:Caux}.
As the strong coupling is running, the integration cannot extend beyond the position of the Landau pole $\Lambda$,
identified by the equation
\beq
1+\as(\mu^2)\beta_0\log\frac{\Lambda^2}{\mu^2} = 0,
\eeq
where $\mu$ is in principle any scale.
Solving the equation, we find that the smallest accessible value of $\xi$ is
\beq
\xi_0 = \frac{\Lambda^2}{Q^2} = \exp\frac{-1}{\beta_0\as(Q^2)} = \frac{\muf^2}{Q^2} \exp\frac{-1}{\beta_0\as(\muf^2)},
\eeq
where we have written $\xi_0$ both in terms of $\mu=Q$ and of $\mu=\muf$.
In particular, since the approximation Eq.~\eqref{eq:UABF} assumes $\as$ to be computed at $\muf$,
the last form is more adequate.
Note that when $\xi=\xi_0$ the approximate evolution factor reduces to
\beq
U(N,Q^2\xi_0,\muf^2) \simeq U_{\rm ABF}\(N,\frac{Q^2}{\muf^2}\xi_0\) = \Big(1-\frac{r(N,\as)}{\beta_0\as}\Big)^{\gamma_+(N,\as)/r(N,\as)},
\eeq
with $\as=\as(\muf^2)$.
This expression is in general finite; however, from general considerations (see Ref.~\cite{Bonvini:2017ogt}),
we expect the evolution function to vanish in $\xi_0$, at least at LL.
The vanishing of $U$ in $\xi_0$ is a property which turns out to be particularly useful,
especially from a numerical point of view.
Thus, to force the evolution function to vanish in $\xi=\xi_0$,
a non-perturbative higher-twist damping function was introduced in Ref.~\cite{Bonvini:2017ogt},
\beq
D_\text{higher-twist}(\xi) =
\begin{cases}
\[1-\(-\as\beta_0\log\xi\)^{1+\frac1{\as\beta_0}}\]\qquad & \xi<1 \\
1 & \xi>1,
\end{cases}
\eeq
such that the final approximated expression for the evolution function is
\begin{align} \label{eq:UABFht}
U(N,Q^2\xi,\muf^2) &\simeq U_{\rm ABF}^{\rm ht}\(N,\frac{Q^2}{\muf^2}\xi\) 
\equiv D_\text{higher-twist}\(\frac{Q^2}{\muf^2}\xi\)\,  U_{\rm ABF}\(N,\frac{Q^2}{\muf^2}\xi\).
\end{align}
This expression, used throughout this paper, also allows to integrate by parts Eqs.~\eqref{eq:CofcalC}
without producing any boundary term.

Finally, we recall that in Ref.~\cite{Bonvini:2016wki} a dedicated anomalous dimension, denoted LL$^\prime$,
was constructed specifically for its usage in the evolution function $U$.
This LL$^\prime$ anomalous dimension is essentially a LL anomalous dimension,
but its dominant small-$x$ singularity is the one of the NLL result.
However, in the recent work of Ref.~\cite{Bonvini:2018xvt}
it has been suggested that this hybrid anomalous dimension may give rise to instabilities when expanded in powers
of $\as$, as needed for the matching of resummed results to fixed order
\new{(we will comment on this in Sect.~\ref{sec:orders})}.
Since the numerical limitations that led to the introduction of the LL$^\prime$ anomalous dimensions
have been overcome in Ref.~\cite{Bonvini:2017ogt},
it has thus been proposed in Ref.~\cite{Bonvini:2018xvt} to use directly the full NLL anomalous dimension,
which also corresponds to the original approach of Ref.~\cite{Altarelli:2008aj}.
In this work we will consider both options later in Sect.~\ref{sec:ggH},
and we will provide further support to the suggestion of Ref.~\cite{Bonvini:2018xvt}
of using the NLL anomalous dimension in the evolution function $U$.
Thus, the new release of \texttt{HELL}, version \texttt{3.0},
performs the resummation using the NLL anomalous dimension in $U$ as default.

To conclude, we report the actual expressions that we will use for the resummation of coefficient functions,
as implemented in the code \texttt{HELL}.
On top of using the approximated evolution function Eq.~\eqref{eq:UABFht},
we integrate by parts so that the derivatives act on the off-shell coefficient function,
and we compute the latter in $N=0$, as its $N$ dependence is subleading.
The results are
\begin{subequations} \label{eq:Cggaux}
\begin{align}
  C_{gg}\(N,\as,\frac{\muf^2}{Q^2}\)
  &= \int_{\xi_0}^\infty d\xi_1\, d\xi_2\; \frac{\partial^2{\cal C}\(0, \xi_1, \xi_2,\as\)}{\partial\xi_1 \partial\xi_2} \, U_{\rm ABF}^{\rm ht}\(N,\frac{Q^2}{\muf^2}\xi_1\) \, U_{\rm ABF}^{\rm ht}\(N,\frac{Q^2}{\muf^2}\xi_2\), \label{eq:Cggaux1}\\
  C_{\rm aux}\(N,\as,\frac{\muf^2}{Q^2}\)
  &= \int_{\xi_0}^\infty d\xi\; \frac{\partial{\cal C}\(0, \xi, 0,\as\)}{\partial\xi} \, U_{\rm ABF}^{\rm ht}\(N,\frac{Q^2}{\muf^2}\xi\). \label{eq:Cggaux2}
\end{align}
\end{subequations}
The second expression is equivalent to the result in the case of a single hadron in the initial state, such as DIS.
The first equation is a new result.
The actual numerical implementation of the first equation further requires (for numerical stability) a change of variables,
as discussed in App.~\ref{sec:IFchangeofvariables}.

\subsection{Equivalence to the impact factor formulation}
\label{sec:equivalence}

In this section, we show that Eq.~\eqref{eq:Cggaux1} leads formally to the same results as the formulation of Ref.~\cite{Ball:2007ra}.
The argument follows closely the one given in Sect.~3.3 of Ref.~\cite{Bonvini:2016wki}, extending it to the case of two initial gluons.
We first introduce the so-called impact factor
\beq\label{eq:impactfactor}
\tilde {\cal C}\(N,M_1,M_2,\as,\frac{\muf^2}{Q^2}\) = \int d\xi_1 \, d\xi_2\; \xi_1^{M_1}\, \xi_2^{M_2}\, \frac{\partial^2{\cal C}\(N, \xi_1, \xi_2,\as\)}{\partial\xi_1 \partial\xi_2} \(\frac{Q^2}{\muf^2}\)^{M_1+M_2},
\eeq
which is simply the double Mellin transform with respect to each $\kt^2$ of the off-shell coefficient function.
For later convenience, we have introduced in the definition of the impact factor a $\muf$-dependent term.
Because by assumptions there are no collinear singularities, this function is analytic in $M_{1,2}=0$, and thus admits an expansion
\beq\label{eq:impactfactorser}
\tilde {\cal C}\(N,M_1,M_2,\as,\frac{\muf^2}{Q^2}\) = \sum_{k,j\geq0} \tilde {\cal C}_{kj}\(N,\as,\frac{\muf^2}{Q^2}\)\, M_1^k\, M_2^j.
\eeq
Because of the symmetry of the off-shell cross section for the exchange of the virtualities, the coefficients of this expansion
are symmetric for the exchange of the indices, $\tilde {\cal C}_{kj} = \tilde {\cal C}_{jk}$.
Now, we write again the off-shell cross section as the double inverse Mellin transform of Eq.~\eqref{eq:impactfactor}, expanded as in Eq.~\eqref{eq:impactfactorser},
\begin{align}\label{eq:d2Cser}
\frac{\partial^2{\cal C}\(N, \xi_1, \xi_2,\as\)}{\partial\xi_1 \partial\xi_2}
&= \int \frac{dM_1}{2\pi i}\, \frac{dM_2}{2\pi i}\; \(\frac{Q^2}{\muf^2}\xi_1\)^{-M_1}\, \(\frac{Q^2}{\muf^2}\xi_2\)^{-M_2}\, \tilde {\cal C}(N,M_1,M_2,\as) \\
&= \sum_{k,j\geq0} \tilde {\cal C}_{kj}\(N,\as,\frac{\muf^2}{Q^2}\) \[\partial_\nu^k\delta\(\nu-\log\(\frac{Q^2}{\muf^2}\xi_1\)\)\]_{\nu=0} \[\partial_\nu^j\delta\(\nu-\log\(\frac{Q^2}{\muf^2}\xi_2\)\)\]_{\nu=0}, \nonumber
\end{align}
where we have used the identity
\beq
\int \frac{dM}{2\pi i}\; \xi^{-M} \, M^k = \[\partial_\nu^k\delta(\nu-\log\xi)\]_{\nu=0}, \qquad k\geq0 .
\eeq
We can now plug Eq.~\eqref{eq:d2Cser} into Eq.~\eqref{eq:Cggaux1} and get\footnote
{Note that the $\muf$ dependence is fully included in the coefficients of the expansion.
If we hadn't included the $\muf$-dependent term in Eq.~\eqref{eq:impactfactor},
then the $\muf$ dependence would be contained in the evolution functions.}
\beq
C_{gg}\(N,\as,\frac{\muf^2}{Q^2}\) = \sum_{k,j\geq0} \tilde {\cal C}_{kj}\(0,\as,\frac{\muf^2}{Q^2}\) \[\partial_\nu^k U_{\rm ABF}^{\rm ht}\(N,e^\nu\)\]_{\nu=0} \[\partial_\nu^j U_{\rm ABF}^{\rm ht}\(N,e^\nu\)\]_{\nu=0}.
\eeq
This expression, computed at central scale $\muf=Q$, reproduces exactly the result of Ref.~\cite{Ball:2007ra}.
Indeed, the derivatives of the evolution function, due to its form Eq.~\eqref{eq:UABF}, satisfy the recursion\footnote
{Note that the higher-twist term does not play any role in this expansion.}
\beq
\[\partial_\nu^{k+1} U_{\rm ABF}^{\rm ht}\(N,e^\nu\)\]_{\nu=0} = \Big(\gamma_+(N,\as) - k\, r(N,\as)\Big) \[\partial_\nu^k U_{\rm ABF}^{\rm ht}\(N,e^\nu\)\]_{\nu=0},
\eeq
which, together with the initial $k=0$ condition $\[U_{\rm ABF}^{\rm ht}\(N,e^\nu\)\]_{\nu=0} = 1$,
give rise to what is sometimes denoted $\[\gamma_+^k\]$ with squared brakets~\cite{Altarelli:2008aj,Bonvini:2016wki,Ball:2013bra}.
In this notation the resummed result is written as
\beq\label{eq:CggresOld}
C_{gg}\(N,\as,\frac{\muf^2}{Q^2}\) = \sum_{k,j\geq0} \tilde {\cal C}_{kj}\(0,\as,\frac{\muf^2}{Q^2}\) \, \[\gamma_+^k(N,\as)\] \, \[\gamma_+^j(N,\as)\],
\eeq
which is a straightforward extension of the analogous resummation in the single-hadron case.
Note that while using Eq.~\eqref{eq:CggresOld} is numerically challenging and necessarily approximate (the infinite series cannot be treated exactly in a numerical code),
and its implementation cannot compete with the straightforward integral representation Eq.~\eqref{eq:Cggaux1},
this form is quite useful for the expansion of the resummed result to fixed order, as we shall now see.

\subsection{Expansion and matching to fixed order}
\label{sec:CFexpansion}

The resummed results Eqs.~\eqref{eq:Cggaux}, which contains the leading small-$x$ contributions to all orders,
are usually matched to a fixed-order contribution. To do so, we need to subtract from the resummed result its expansion in $\as$
up to the fixed-order $k$ considered,
\beq
\Delta_k C\(N,\as,\frac{\muf^2}{Q^2}\) = C\(N,\as,\frac{\muf^2}{Q^2}\) - \sum_{j=0}^k \as^j \, C^{(j)}\(N,\frac{\muf^2}{Q^2}\),
\eeq
where the last sum is the truncated $\as$-expansion of the first (resummed) coefficient $C$.
Then, this $\Delta_kC$ contribution is of $\Ord(\as^{k+1})$, and can be safely added to the fixed N$^k$LO result.
In this work, we consider the matching up to N$^3$LO, which is the highest fixed-order accuracy
available for Higgs production in gluon fusion.
Thus, we need the expansion of the resummation up to $\Ord(\as^3)$.

The construction of this expansion is obtained in a simple way using the impact-factor formulation, Eq.~\eqref{eq:CggresOld}.
To use it, we first write explicitly $\[\gamma_+^k\]$ up to $k=3$ (omitting the arguments for ease of notation),
\begin{align}\label{eq:gammaSB}
\[\gamma_+^0\] &= 1, \nonumber\\
\[\gamma_+^1\] &= \gamma_+, \nonumber\\
\[\gamma_+^2\] &= \gamma_+\(\gamma_+ - r\), \nonumber\\
\[\gamma_+^3\] &= \gamma_+\(\gamma_+ - r\) \(\gamma_+ - 2r\),
\end{align}
where $r$ is given in Eq.~\eqref{eq:gammaasapprox}.
To proceed, we now need to expand in powers of $\as$ both $\gamma_+$ and $r$.
However, before doing so, we recall that in Refs.~\cite{Bonvini:2016wki,Bonvini:2017ogt} a variant of the resummation,
used to estimate the uncertainty from subleading contributions, was introduced in which $r$ is replaced with $\as\beta_0$,
i.e.\ the $\as$-dependence of the anomalous dimension is treated as if it was just $\Ord(\as)$, in line with the approximation Eq.~\eqref{eq:gammaasapprox}.
To cover both cases, up to N$^3$LO it is sufficient to introduce a single parameter $T$, which equals 2 in the default case, and equals 1 in the limit $r=\as\beta_0$.
Introducing the expansion of the anomalous dimension
\beq
\gamma_+ = \as \gamma_0 + \as^2 \gamma_1 + \as^3 \gamma_2 + \Ord(\as^4)
\eeq
we can indeed write
\beq
r = \as\beta_0 \[1+(T-1)\as\frac{\gamma_1}{\gamma_0}+\Ord(\as^2)\].
\eeq
With these expressions we can expand Eq.~\eqref{eq:gammaSB} as
\begin{align}\label{eq:gammaSB2}
\[\gamma_+^0\] &= 1, \nonumber\\
\[\gamma_+^1\] &= \as \gamma_0 + \as^2 \gamma_1 + \as^3 \gamma_2 + \Ord(\as^4), \nonumber\\
\[\gamma_+^2\] &= \as^2 \gamma_0\(\gamma_0-\beta_0\) + \as^3 \gamma_1 \(2\gamma_0-T\beta_0\) +\Ord(\as^4), \nonumber\\
\[\gamma_+^3\] &= \as^3 \gamma_0\(\gamma_0-\beta_0\)\(\gamma_0-2\beta_0\) +\Ord(\as^4),
\end{align}
which can be now used in Eq.~\eqref{eq:CggresOld} to get the $\as$-expansion of the $gg$ coefficient function:
\begin{align}\label{eq:Cggexp}
C_{gg} &= \tilde{\cal C}_{00} +\as 2\tilde {\cal C}_{10}\gamma_0 + \as^2\[2\tilde {\cal C}_{10}\gamma_1+2\tilde {\cal C}_{20}\gamma_0\(\gamma_0-\beta_0\)+\tilde {\cal C}_{11}\gamma_0^2\] \nonumber\\
& + \as^3\[2\tilde {\cal C}_{10}\gamma_2+2\tilde {\cal C}_{20}\gamma_1\(2\gamma_0-T\beta_0\)+2\tilde {\cal C}_{11}\gamma_0\gamma_1+2\tilde {\cal C}_{30}\gamma_0\(\gamma_0-\beta_0\)\(\gamma_0-2\beta_0\)
+2\tilde {\cal C}_{21}\gamma_0^2\(\gamma_0-\beta_0\)\] \nonumber\\
&+\Ord(\as^4).
\end{align}
Depending on the anomalous dimension used in the evolution function $U$ (see discussion in Sect.~\ref{sec:evolU}),
which determines the actual form of $\gamma_{0,1,2}$,
this expression provides the first few orders of the resummed coefficient function
needed to construct the resummed contribution $\Delta_kC_{gg}$ up to $k=3$.

To construct the expansion of the resummed coefficient functions for the other partonic channels,
we need to expand the auxiliary function Eq.~\eqref{eq:Cggaux2}.
Straightforwardly, its impact-factor form can be derived from Eq.~\eqref{eq:CggresOld} by keeping only the $j=0$ part of the sum,
and flipping the sign
\beq\label{eq:CauxresOld}
C_{\rm aux}\(N,\as,\frac{\muf^2}{Q^2}\) = -\sum_{k\geq0} \tilde {\cal C}_{k0}\(0,\as,\frac{\muf^2}{Q^2}\) \, \[\gamma_+^k(N,\as)\].
\eeq
Thus, its $\as$ expansion is given by
\begin{align}\label{eq:Cauxexp}
C_{\rm aux} &= -\tilde{\cal C}_{00} -\as \tilde {\cal C}_{10}\gamma_0 - \as^2\[\tilde {\cal C}_{10}\gamma_1+\tilde {\cal C}_{20}\gamma_0\(\gamma_0-\beta_0\)\] \nonumber\\
& - \as^3\[\tilde {\cal C}_{10}\gamma_2+\tilde {\cal C}_{20}\gamma_1\(2\gamma_0-T\beta_0\)+\tilde {\cal C}_{30}\gamma_0\(\gamma_0-\beta_0\)\(\gamma_0-2\beta_0\)\] \nonumber\\
&+\Ord(\as^4).
\end{align}
Form Eqs.~\eqref{eq:Cggexp} and \eqref{eq:Cauxexp} we can construct the expansions of the quark coefficient functions, according to Eqs.~\eqref{eq:CofCggaux},
\begin{align}
C_{qg} &= \frac{C_F}{C_A}\bigg[\as \tilde {\cal C}_{10}\gamma_0 + \as^2\[\tilde {\cal C}_{10}\gamma_1+\tilde {\cal C}_{20}\gamma_0\(\gamma_0-\beta_0\)+\tilde {\cal C}_{11}\gamma_0^2\] \nonumber\\
&\qquad + \as^3\[\tilde {\cal C}_{10}\gamma_2+\tilde {\cal C}_{20}\gamma_1\(2\gamma_0-T\beta_0\)+2\tilde {\cal C}_{11}\gamma_0\gamma_1
  +\tilde {\cal C}_{30}\gamma_0\(\gamma_0-\beta_0\)\(\gamma_0-2\beta_0\)
  +2\tilde {\cal C}_{21}\gamma_0^2\(\gamma_0-\beta_0\)\] \nonumber\\
&\qquad+\Ord(\as^4) \bigg],\label{eq:Cqgexp}\\
C_{qq} &= \(\frac{C_F}{C_A}\)^2\bigg[\as^2\tilde {\cal C}_{11}\gamma_0^2 + \as^3\[2\tilde {\cal C}_{11}\gamma_0\gamma_1 +2\tilde {\cal C}_{21}\gamma_0^2\(\gamma_0-\beta_0\)\]+\Ord(\as^4)\bigg].\label{eq:Cqqexp}
\end{align}
With these expressions it is then possible to construct also the resummed contributions
$\Delta_kC_{qg}$ and $\Delta_kC_{qq}$ for the quark coefficient functions up to $k=3$.
All together, these expressions allow to match resummed results to N$^3$LO.
The computation of the $\tilde{\cal C}_{ij}$ coefficients, needed for the expansions presented here,
is detailed in App.~\ref{sec:IFcoeff}.

\subsection{The first few orders of the anomalous dimension at LL$^\prime$ and NLL}
\label{sec:orders}

To conclude the section, we now present the analytic expressions of the 
$\Ord(\as^{1,2,3})$ anomalous dimensions $\gamma_{0,1,2}$ needed for the matching of
the resummed coefficient function to fixed order up to N$^3$LO, Eqs.~\eqref{eq:Cggexp}, \eqref{eq:Cqgexp} and \eqref{eq:Cqqexp}.
We treat both the case in which the anomalous dimension used is the LL$^\prime$ introduced in Ref.~\cite{Bonvini:2016wki}
and the case in which the full NLL anomalous dimension is used,
as suggested in Ref.~\cite{Bonvini:2018xvt,Altarelli:2008aj}, see discussion in Sect.~\ref{sec:evolU}.
These expressions are obtained by expanding the purely resummed LL$^\prime$ or NLL anomalous dimension,
and have been already computed and presented in Refs.~\cite{Bonvini:2016wki,Bonvini:2017ogt,Bonvini:2018xvt}.
Thus, here we only report the final results~\cite{Bonvini:2018xvt}.
For LL$^\prime$ resummation we have
\begin{align}
\gamma_0^{\rm LL'} &= \frac{a_{11}}{N} + \frac{a_{10}}{N+1}, \label{eq:gamma0LLp}\\
\gamma_1^{\rm LL'} &= \beta_0a_{11}\(\frac{21}{8}\zeta_3-4\log2\)\(\frac1N-\frac{4N}{(N+1)^2}\) \label{eq:gamma1LLp}\\
\gamma_2^{\rm LL'} &= \frac{\lambda_2}{N^2}+\frac{\lambda_1}{N} - \(\lambda_2+\lambda_1\) \frac{4N}{(N+1)^2} \nonumber\\
                   &\quad+\(\frac{a_{11}}{N^2}+\frac{2(a_{11}+a_{10})}{(N+1)^2}\)
                     \bigg[ \frac{a_{11}a_{10}}{(1+N)^2}-\frac{a_{11}a_{10}}{4} \frac{4N}{(N+1)^2} \nonumber\\
&\qquad \qquad \qquad \qquad \qquad \qquad\; +a_{11}\(\frac{a_{11}}N+a_{10}-\frac{2(a_{11}+a_{10})N}{N+1}\)\[\psi_1(1+N)-\zeta_2\]\bigg],
\label{eq:gamma2LLp}
\end{align}
while for NLL resummation the results are
\begin{align}
\gamma_0^{\rm NLL} &= \frac{a_{11}}{N} + \frac{a_{10}}{N+1}, \label{eq:gamma0NLL}\\
\gamma_1^{\rm NLL} &= \frac{a_{21}}{N} - \frac{2a_{21}}{N+1}, \label{eq:gamma1NLL}\\
\gamma_2^{\rm NLL} &= \frac74\beta_0^2a_{11}\zeta_3 \(\frac1N - \frac{4N}{(N+1)^2}\) \nonumber\\
                   &\quad+\(\frac{a_{11}}{N^2}+\frac{2(a_{11}+a_{10})}{(N+1)^2}\)
                     \bigg[ \rho + \frac{a_{21}}{1+N} + \frac{a_{11}a_{10}}{(1+N)^2} -\(\rho + \frac{a_{21}}2 + \frac{a_{11}a_{10}}4 -\beta_0a_{11}\) \frac{4N}{(N+1)^2} \nonumber\\
&\qquad \qquad \qquad \qquad \qquad \qquad\; +a_{11}\(\frac{a_{11}}N+a_{10}-\frac{2(a_{11}+a_{10})N}{N+1}+\beta_0\)\[\psi_1(1+N)-\zeta_2\]\bigg].
\label{eq:gamma2NLL}
\end{align}
The coefficients appearing above are
\begin{subequations}\label{eq:aij}
\begin{align}
  a_{11} &= \frac{C_A}\pi, \\
  a_{21} &= n_f\frac{26C_F-23C_A}{36\pi^2}, \\
  a_{10} &= -\frac{11C_A + 2n_f(1-2C_F/C_A)}{12\pi},
\end{align}
\end{subequations}
and
\begin{subequations}\label{eq:rhos}
\begin{align}
\lambda_2 &= 1.26717 + 0.110072 n_f,\\
\lambda_1 &=
            \begin{cases}
              -60.6782 + 3.53857 n_f + 0.00841828 n_f^2 &\quad \text{(default)}\\
              -30.3568 + 1.77143 n_f + 0.00414421 n_f^2 &\quad \text{(variant)},
            \end{cases}
\label{eq:lambda1}\\
  \rho &= \frac{1}{\pi^2} \[ C_A^2\(-\frac{74}{27}+\frac{11}{12}\zeta_2+\frac52\zeta_3\)
         + n_fC_A\(\frac4{27}+\frac16\zeta_2\) + n_fC_F\(\frac7{27}-\frac13\zeta_2\) \].
\end{align}
\end{subequations}
The two values of $\lambda_1$, Eq.~\eqref{eq:lambda1}, come from another variant of the resummation,
used in the construction of $\gamma_+$, which affects only the LL$'$ anomalous dimension at this order.
More details can be found in App.~A of Ref.~\cite{Bonvini:2018xvt}.
All these expressions are implemented in \texttt{HELL~3.0}.

\new{Before moving on, we would like to comment on a particular feature of these expansions.
We recall that, due to accidental cancellations, the expected leading singularities of the NLO and NNLO anomalous dimensions
are zero. Since both LL$'$ and NLL anomalous dimension are accurate at LL,
the leading terms $1/N^2$ in $\gamma_1$ and $1/N^3$ in $\gamma_2$ are correctly absent.
As a consequence, the highest singularity in $\gamma_1$ and $\gamma_2$ is the NLL one, which is correct
only in the NLL anomalous dimension.
Instead, the dominant singularity (and any other subleading term) of these two orders in the LL$'$ result is not correct.
Thus, while the all-order LL$'$ and NLL anomalous dimensions may be in good agreement (and indeed they are),
their $\Ord(\as^2)$ and $\Ord(\as^3)$ expansions may be very different (and indeed they differ substantially at $\Ord(\as^3)$).
For this reason, resummed results which depend explicitly on $\gamma_1^{\rm LL'}$ and $\gamma_2^{\rm LL'}$
(such as resummed results matched to NNLO and beyond) may differ substantially from those obtained with the NLL anomalous dimension,
and when this is the case results obtained in the NLL case have to be favoured.
It has been noticed in Ref.~\cite{Bonvini:2018xvt} (and we will see it also here in Sect.~\ref{sec:ggH})
that when matching to N$^3$LO resummed results based on LL$'$ behave pathologically at medium/large values of $x$,
which is a consequence of a similar behaviour in the inverse Mellin transform of $\gamma_2^{\rm LL'}$.
This is the main motivation that induced Ref.~\cite{Bonvini:2018xvt} to propose the use of the NLL anomalous dimension as default.}

\section{Resummed Higgs cross section at the LHC and beyond}
\label{sec:ggH}

We now turn our attention to a hadron-hadron collider process which is of great interest for LHC phenomenology: Higgs production in gluon fusion
($ggH$ for short).
Of course, Higgs physics is very interesting because the Higgs sector can be sensitive to new physics beyond the Standard Model.
The inclusive Higgs cross section, which we are going to consider, is for instance sensitive to heavy particles
coupling to gluons, which may then run in the loop of Fig.~\ref{fig:ggHprocLO} and alter the production rate at the LHC.

Moreover, from a theoretical point of view, Higgs production is an interesting process because
fixed-order perturbative QCD corrections are very large, with NLO being about twice the LO, and NNLO adding another $\sim40\%$
of the LO to the cross section.
This motivated various studies to go beyond NNLO~\cite
{Moch:2005ky,Ahrens:2008qu,Ball:2013bra,Anastasiou:2014vaa,Bonvini:2014jma,deFlorian:2014vta,Anastasiou:2014lda},
culminating in the computation of this production process to N$^3$LO~\cite{Anastasiou:2015ema,Anzai:2015wma,Anastasiou:2016cez,Mistlberger:2018etf}
in the large top-mass limit.\footnote
{A consistent N$^3$LO calculation would also require PDFs fitted and evolved with N$^3$LO theory.
However, the four-loop DGLAP splitting functions are not yet fully known,
though recently there has been impressive progress towards their computation~\cite{Davies:2016jie,Moch:2017uml,Vogt:2018ytw}.
Thus, at the moment all N$^3$LO predictions use NNLO PDFs.}
It has been demonstrated in various ways that such large corrections mostly originate
from soft-virtual contributions~\cite
{Ahrens:2008qu,Bonvini:2012an,Ball:2013bra,deFlorian:2014vta,Anastasiou:2014lda}, dominant at large $x$,
and can be resummed to all orders by means of threshold resummation techniques~\cite
{Catani:2003zt,Moch:2005ky,Ahrens:2008nc,deFlorian:2012yg},
reaching N$^3$LL accuracy~\cite{Bonvini:2014joa,Catani:2014uta,Bonvini:2014tea,Schmidt:2015cea,Bonvini:2016frm}.\footnote
{To be precise, all contributions that are relevant for N$^3$LL are
known~\cite{Moch:2005ba,Moch:2005ky,Laenen:2005uz,Anastasiou:2014vaa},
with the exception of the four-loop cusp anomalous dimension
(see~\cite{Boels:2017skl,Grozin:2017css} for recent progress),
which is thought to have a negligible impact.}
N$^3$LO+N$^3$LL predictions are very close to N$^3$LO ones, thus suggesting that perturbative expansion
is apparently converging, and giving some confidence that current theoretical predictions, such as
the one recommended by the LHC Higgs Cross Section Working Group (HXSWG)~\cite{deFlorian:2016spz},
are sufficiently accurate for precision phenomenology.

In this work we investigate the effect of the all-order resummation of the small-$x$ logarithms,
i.e.\ those important in the opposite limit with respect to the one largely studied for this process.
Indeed, Higgs production in gluon fusion is also one of the LHC processes which is expected to be most sensitive to
small-$x$ logarithmic enhancement, due to the fact that it is gluon-gluon initiated at lowest order,
and the gluon PDF is the most sensitive to small-$x$ resummation effects.
We will show that the consistent inclusion of small-$x$ resummation has a sizeable effect.
In the $Q_0\MSbar$ scheme that we adopt, most of this effect comes from the use of resummed PDF
instead of fixed-order ones, while the effect of resummation in the coefficient function is much milder.
The effect of resummation gets progressively larger as the collider energy is increased,
since smaller and smaller values of $x$ become accessible and increasingly important.
Therefore, the inclusion of resummation becomes more crucial at higher-energy colliders,
such as the High-Energy phase of the LHC, with $\sqrt{s}=27$~TeV,
and even more at a Future Circular hadron Collider (FCC-hh) of $\sqrt{s}=100$~TeV.

Before presenting resummed results, we recall that many results in the computation of the Higgs production
cross section in gluon fusion are obtained within the so-called large top-mass effective field theory (EFT henceforth),
where the top-quark is integrated out of the theory
and its effect included as corrections in powers of $\mh^2/\mt^2$.
However, in this theory the small-$x$ region cannot be predicted correctly, as the $x\to0$ limit does not commute
with the $\mt\to\infty$ limit of the EFT.
Therefore, the correct inclusion of small-$x$ resummation also requires a correct treatment of the small-$x$ region
at fixed order.
In Sect.~\ref{sec:ggHfo}, we then first revisit how top mass dependence is included in fixed-order result
and how the correct small-$x$ logarithms can be included at fixed order.
Then, in Sect.~\ref{sec:ggHpartonic} and Sect.~\ref{sec:ggHxs} we will show the impact of small-$x$ resummation
at parton level and on the cross section, respectively.

This section provides a detailed explanation of the small-$x$ resummed results presented in Ref.~\cite{Bonvini:2018ixe}
in the context of a double small-$x$ plus large-$x$ resummation.

\subsection{Construction of the fixed-order result at small $x$ with top mass dependence}
\label{sec:ggHfo}

The LO diagram for $ggH$ production, Fig.~\ref{fig:ggHprocLO}, is a one-loop diagram with a massive internal particle.
The NLO correction to this process has been carried out exactly~\cite{Spira:1995rr,Bonciani:2007ex}.
However, from NNLO onward, the exact computation would require the evaluation of three-loop (or higher) diagrams with massive internal lines,
which are out of reach of the current computational technology.

However, in the limit in which the partonic center-of-mass energy $\sqrt{\hat s}$ is (much) smaller than twice the top mass $\mt$,
one can construct an effective field theory (EFT) in which the top loop shrinks to a point,
leading to a pointlike interaction described by operators. The operator with the lower dimensionality
does not depend explicitly on the top mass, except for a logarithmic dependence appearing in its Wilson coefficient.
Operators with higher dimensionality will give rise to corrections suppressed by increasing powers of $1/\mt^2$.

Within this EFT it has been possible to push the computation of the $ggH$ cross section
at NNLO (both at the leading power level~\cite{Anastasiou:2002yz,Harlander:2002wh,Ravindran:2003um}
and including a few corrections in $1/\mt^2$~\cite{Harlander:2009mq,Harlander:2009my,Pak:2009dg})
and even at N$^3$LO~\cite{Anastasiou:2015ema,Anastasiou:2016cez,Anzai:2015wma,Mistlberger:2018etf} (at leading power).
Because the expansion parameter of the EFT is
\beq\label{eq:EFTexppar}
\frac{\hat s}{4\mt^2} = \frac{\mh^2}{4z\mt^2},
\eeq
where $z=\mh^2/\hat s$ and $\hat s$ the partonic center-of-mass energy,
the limits $\mt\to\infty$ and $z\to0$ do not commute.\footnote
{Note that the variable $z$ is a scaling variable as $x$ in DIS and in the PDFs, and thus small-$x$ resummation
in the Higgs partonic coefficient functions resums logarithms of $z$.}
Thus, the EFT cannot describe the small-$z$ region correctly.
For this reason, computations within the EFT can be (and have been) carried out
as threshold expansions about $z=1$, i.e.\ as power series in $(1-z)$, e.g.\ in Refs.~\cite{Harlander:2009mq,Anastasiou:2016cez}.
Indeed, at large and medium $z$ this expansion converges to the exact result, while at small $z$ it is wrong anyway.

The goal of this subsection is to provide a way to supplement computations in the EFT with the exact small-$z$ logarithms,
which can be predicted from the resummation.
Let's consider the generic coefficient function with perturbative expansion
(omitting all unnecessary arguments and indices for simplicity)
\beq
C\(z, \as, \frac{\mh^2}{\mt^2}\) = \sum_{k=0}^\infty \as^k C^{(k)}\(z,\frac{\mh^2}{\mt^2}\).
\eeq
As we already stated, from NNLO onwards the exact $\mh/\mt$ dependence is unknown.
At NNLO, $\mh/\mt$ effects have been computed as an expansion in
\beq
\rhot = \frac{\mh^2}{\mt^2}
\eeq
up to the order $p_{\rm max}=3$~\cite{Harlander:2009mq,Harlander:2009my,Pak:2009dg},
\begin{align}
  C^{(2)}\(z,\rhot\) 
  &\simeq \sum_{p=0}^{p_{\rm max}} \rhot^p C_p^{(2)}(z),
\end{align}
while at N$^3$LO only the first term is known ($p_{\rm max}=0$)~\cite
{Anastasiou:2015ema,Anastasiou:2016cez,Anzai:2015wma,Mistlberger:2018etf}.
However, the expansion in $\mh/\mt$ is not accurate at small $z$, since the actual expansion parameter is
the one given in Eq.~\eqref{eq:EFTexppar}: in particular, the $\rhot$ expansion is supposed to break down for $z\lesssim\rhot/4$.
Therefore, the small-$z$ behaviour of the $\mh/\mt$ expansion is unstable,
\begin{align}\label{eq:CkpSX}
C^{(k)}(z,\rhot) 
  &= \sum_{p=0}^{p_{\rm max}} \rhot^p \sum_{j=0}^p\sum_{n=0}^{2k-1} B_{p,j,n}^{(k)}\frac{\log^n z}{z^{1+j}} + \Ord(z^0) + \Ord(\rhot^{p_{\rm max}+1}),
\end{align}
exhibiting double-logarithmic enhancement and higher powers of $1/z$ at each extra order in $\rhot$,
in contrast with the exact small-$z$ behaviour
\begin{align}\label{eq:exactSX}
  C^{(k)}\(z,\rhot\) 
  &= \sum_{n=0}^{k-1} A_n^{(k)}(\rhot)\frac{\log^nz}{z} + \Ord(z^0),
\end{align}
which is single-logarithmic enhanced and always contains a single power of $1/z$.
The exact small-$z$ behaviour, Eq.~\eqref{eq:exactSX}, can be predicted (at least at LL, i.e.\ $n=k-1$) from high-energy resummation.
Our goal is therefore to understand how the exact Eq.~\eqref{eq:exactSX} can be used to replace the wrong
Eq.~\eqref{eq:CkpSX} of the large $\mt$ EFT computation.
We recall that two different phenomenological solutions to this problem have been already proposed
in Refs.~\cite{Harlander:2009mq,Harlander:2009my} and \cite{Pak:2009dg}, respectively.
Here, we want to deal with this problem in a systematic way.

As a first step, we need to understand whether the limit $p_{\rm max}\to\infty$ converges to the exact result or not.
At large $z$ and for sufficiently small $\rhot$, the answer must be yes, or at least asymptotically yes.
On the other hand, at small $z$ the expansion clearly diverges, with new singularities appearing at each order in $\rhot$, Eq.~\eqref{eq:CkpSX}.
Thus, at small $z$, only the all-order sum may make sense, but certainly not any finite truncation of it.
Therefore, in order to build up a sensible result, we need to make sure first to get rid of the
bad small-$z$ behaviour of the $\rhot$ expansion, and once this is done we can add the exact small-$z$ contribution, Eq.~\eqref{eq:exactSX}.
The final expression must be such that the limit $p_{\rm max}\to\infty$ tends to the exact result.\footnote
{Possibly up to subleading power logarithmic contributions behaving as $\log z$ without any $1/z$ enhancement.}
We will consider four possible approaches, in turn.

\paragraph{Method of subtraction}

The first option that we consider consists in subtracting from the $\rhot$ expansion the ``wrong'' small-$z$ behaviour,
Eq.~\eqref{eq:CkpSX}, replacing it with the exact small $z$, Eq.~\eqref{eq:exactSX}.
The resulting expression is
\begin{align}\label{eq:Ckbest}
  C^{(k)}\(z,\rhot\)
  &\simeq \sum_{p=0}^{p_{\rm max}} \rhot^p \[C_p^{(k)}(z) - d(z) \sum_{j=0}^p\sum_{n=0}^{2k-1} B_{p,j,n}^{(k)}\frac{\log^n z}{z^{1+j}}  \]
  + d(z)\sum_{n=0}^{k-1} A_n^{(k)}(\rhot)\frac{\log^nz}{z},
\end{align}
where we have further introduced a function $d(z)$,
which represents a possible large-$z$ damping to be uniformly applied to the small-$z$ parts of the result.
The role of this damping is to suppress the effect of the small-$z$ contributions at large $z$:
indeed, the $1/z$ terms without logarithms contained in the small-$z$ contributions do not vanish at large $z$.

This method, despite its simplicity and naturalness, has \new{three } important drawbacks.
The first is that it requires the exact EFT result, and not just its (simpler to compute) threshold expansion,
which, as we already mentioned, carries the same correct information, and differs only in the region where they are both wrong.
At NNLO, the EFT small-$z$ contribution is fully known for $p=0,1,2$~\cite{Pak:2009dg},
while the threshold expansion is also known for $p=3$~\cite{Harlander:2009mq,Harlander:2009my}.
At N$^3$LO, only the leading term $B^{(3)}_{0,0,5}$ was known until very recently,
when in Ref.~\cite{Mistlberger:2018etf} the exact leading EFT result ($p=0$) has been computed,
thus allowing to use this method at N$^3$LO as well.
The second
\new{
is that the function in squared brackets in Eq.~\eqref{eq:Ckbest} still contains
double-logarithmic terms at $\Ord(z^0)$ which are not predicted correctly by the EFT expansion either,
and can thus potentially contaminate the result.
(In principle these logarithmic contributions could be subtracted as well,
however their counterparts in the exact theory cannot be derived from small-$z$ resummation and thus they cannot be added back.)
The third
}
and perhaps more severe issue is that the result may strongly depend on the damping function used.
Indeed, ideally, the two small-$z$ contributions should cancel each other at large $z$.
However, since the $z\to1$ limit of the small-$z$ contribution in the $\rhot$ expansion inherits its instability,
there is no practical compensation at large $z$ between what is subtracted and what is added.
And this must not happen, since the $C^{(k)}_p(z)$ terms are supposed to be reliable in the $z\to1$ limit.
Thus the damping becomes a necessity, but its form is not prescribed by the procedure,
leaving a degree of arbitrariness which may contaminate the result.

\paragraph{Method of threshold expansion}

The expression in square brakets in Eq.~\eqref{eq:Ckbest} does no longer contain divergent terms at small $z$.\footnote
{Except for the aforementioned powers of $\log z$ without $1/z$ enhancement, which are not predicted correctly either.}
Thus, there is no loss of information if it is replaced with its threshold expansion, i.e.\ an expansion in powers of $(1-z)$.
Let us define
\beq\label{eq:DeltaCdef}
\delta C_p^{(k)}(z) = C_p^{(k)}(z) - d(z) \sum_{j=0}^p\sum_{n=0}^{2k-1} B_{p,j,n}^{(k)}\frac{\log^n z}{z^{1+j}}
\eeq
to be the function in square brackets in Eq.~\eqref{eq:Ckbest}.
Eq.~\eqref{eq:DeltaCdef} can be expanded at large $z$ as\footnote
{\label{foot:reg}To simplify the following discussion, let us assume that for the $gg$ channel the coefficient function
is defined as the ``regular'' part of the decomposition
$
C^{(k)}_{gg} = \big[C^{(k)}_{gg}\big]_{\rm distr} + \big[C^{(k)}_{gg}\big]_{\rm reg},
$
where the distributional part contains plus distributions and $\delta(1-z)$ functions, and the regular part everything else.}
\begin{align}
\delta C_p^{(k)}(z) 
  & = z^a \[\frac{\delta C_p^{(k)}(z)}{z^a}\]_{\rm t.e.} \nonumber\\
  & = z^a\sum_{i=0}^\infty (1-z)^i \[ c_{p,i}^{(k)}(a,\ell) - \sum_{j=0}^p\sum_{n=0}^{2k-1} B_{p,j,n}^{(k)}b_{j,n,i}(a)\],
\label{eq:DeltaCkse}
\end{align}
where $a$ is a parameter, and the subscript ``t.e.''\ (threshold expansion) means that the function enclosed by those brackets
is expanded in powers of $1-z$.
In the equation above, the expansion coefficients $c_{p,i}^{(k)}$ also depend in general on
\beq
\ell\equiv \log(1-z),
\eeq
which is clearly not expandable in $z=1$, and we have introduced
the coefficients $b_{j,n,i}(a)$ according to
\beq\label{eq:logexp}
d(z)\frac{\log^n z}{z^{1+j+a}} = \sum_{i=0}^\infty (1-z)^i b_{j,n,i}(a),
\eeq
which thus depend on the damping function $d(z)$.
The $a$ parameter is in principle free, since the result is independent of $a$ when the whole series in $1-z$ is considered.
Of course, any finite truncation of the series to $i=i_{\rm max}$ will have a residual dependence on $a$,
which can thus be used e.g.\ to estimate the uncertainty due to missing terms in the threshold expansion~\cite{Anastasiou:2016cez}.
The coefficients $C_p^{(k)}(z)$ have been computed in the first place as a threshold expansion
at NNLO~\cite{Harlander:2009mq,Harlander:2009my} and N$^3$LO~\cite{Anastasiou:2016cez},
so the relevant $c_{p,i}^{(k)}$ are all known.

Let us comment on the choice of the parameter $a$.
The value $a=-1$ is the one adopted in Ref.~\cite{Harlander:2009mq,Harlander:2009my}
(as there the partonic cross section $zC(z)$ is expanded).\footnote
{To be precise, in Ref.~\cite{Harlander:2009mq,Harlander:2009my} also the distributional part in the $gg$ channel is multiplied by $1/z$, thus changing the actual $c_{p,i}^{(k)}$ coefficients. However, this difference is immaterial for the present discussion.}
This choice is such that terms behaving as $1/z$ are generated in the threshold expansion;
however, these terms are not predicted correctly by the EFT, and indeed they have been subtracted in Eq.~\eqref{eq:DeltaCdef},
so producing them can be dangerous.
On the contrary, we note that if we choose $a\geq 0$ both terms in Eq.~\eqref{eq:DeltaCdef} lead to a threshold expansion
which does not grow at small $z$.
In particular, for $a=0$ the threshold expansion goes to a constant, while for larger $a$ it vanishes in $z=0$
(however $a$ cannot be too large, otherwise it would affect the coefficient function in a region of medium $z$ where the theshold expansion is reliable).
This means that choosing $a\geq0$ the resulting coefficient function does not contain any leading small-$z$ contribution.
Thus, the threshold expansion with $a\geq0$ provides a natural and legitimate way of damping the EFT result at small-$z$,
\new{thereby also removing the potential danger induced by the EFT logarithmic terms at $\Ord(z^0)$.}

This observation may suggest that, as long as the coefficient function is threshold-expanded with $a\geq0$,
the term $\delta C^{(k)}_p(z)$, Eq.~\eqref{eq:DeltaCdef}, appearing in Eq.~\eqref{eq:Ckbest}
can be replaced with just the full coefficient function $C^{(k)}_p(z)$,
without subtracting the small-$z$ terms. Indeed, at large $z$ the two objects do not differ,
due to the damping $d(z)$ which suppresses the subtraction terms,
and at small $z$ both objects do not contain small-$z$ contributions.
Clearly, the subleading small-$z$ terms (those not enhanced by $1/z$) may differ, but these are anyway beyond our control,
and certainly not predicted by the last term of Eq.~\eqref{eq:Ckbest}.
Thus, we may conclude that an equally good definition of the full result is
\begin{align}\label{eq:Ckbest2}
  C^{(k)}\(z,\rhot\)
  &\simeq z^a\sum_{p=0}^{p_{\rm max}} \rhot^p \sum_{i=0}^{i_{\rm max}} (1-z)^i c_{p,i}^{(k)}(a,\ell)
    + d(z)\sum_{n=0}^{k-1} A_n^{(k)}(\rhot)\frac{\log^nz}{z},
\end{align}
provided $a\geq0$.
In fact, Eq.~\eqref{eq:Ckbest2} can be obtained with no approximations, by exploiting
the dependence on the damping function $d(z)$.
Indeed, in this case, the damping function is no longer fully free,
but we have a guiding principle how to choose its form.
Namely, since at large $z$ the first part of Eq.~\eqref{eq:Ckbest2}
is reliable up to $\Ord\((1-z)^{i_{\rm max}}\)$, the damping function must be suppressed at least as
\beq\label{eq:damping1}
d(z) = (1-z)^{i_{\rm max}+1},
\eeq
such that the exact small-$z$ part does not spoil the accuracy of the threshold expansion.
With this choice for $d(z)$, the $b_{j,n,i}(a)$ coefficients are all zero for $i\leq i_{\rm max}$:
hence, up to $i=i_{\rm max}$, the EFT small-$z$ terms \emph{do vanish}.
Thus, when using this damping function (or a more suppressed version of it), the threshold expansion of Eq.~\eqref{eq:Ckbest}
gives \emph{exactly} Eq.~\eqref{eq:Ckbest2}.
Note that, because in Eq.~\eqref{eq:Ckbest2} there is no subtraction of small-$z$ EFT contributions,
this formulation is simpler and very suitable for numerical implementation, both at NNLO and at N$^3$LO.

\paragraph{Method of double subtraction}

In Ref.~\cite{Harlander:2009mq,Harlander:2009my} a different construction was considered,
where the exact small $z$ is added to the threshold expansion of $C^{(k)}_p(z)$ after having subtracted from it its own threshold expansion,
without applying any damping.
To derive this possible approach, let us start again from Eq.~\eqref{eq:Ckbest}, to which we replace the first part with its threshold expansion,
and where we add and subtract the threshold expansion of the exact small $z$,
\begin{align}\label{eq:Ckbest3}
  C^{(k)}\(z,\rhot\)
  &\simeq z^a\sum_{i=0}^{i_{\rm max}}(1-z)^i \[\sum_{p=0}^{p_{\rm max}} \rhot^p \(c_{p,i}^{(k)}(a,\ell) - \sum_{j=0}^p\sum_{n=0}^{2k-1} B_{p,j,n}^{(k)}b_{j,n,i}(a)\) +\sum_{n=0}^{k-1} A_n^{(k)}(\rhot) b_{0,n,i}(a) \]
    \nonumber\\
  &\quad+ \sum_{n=0}^{k-1} A_n^{(k)}(\rhot)\[d(z)\frac{\log^nz}{z} - z^a\sum_{i=0}^{i_{\rm max}}(1-z)^i b_{0,n,i}(a) \].
\end{align}
As far as the exact small-$z$ term is concerned, it is clear that the damping function becomes unnecessary with this approach,
as the term in square brackets in the last line of Eq.~\eqref{eq:Ckbest3} is of $\Ord\((1-z)^{i_{\rm max}+1}\)$ irrespectively of the form of $d(z)$.
If we choose the damping function as in Eq.~\eqref{eq:damping1} then we recover exactly Eq.~\eqref{eq:Ckbest2},
since $b_{j,n,i}(a)=0$ for $i\leq i_{\rm max}$.
However, following Ref.~\cite{Harlander:2009mq,Harlander:2009my}, we can now choose
\beq
d(z)=1,
\eeq
thus fixing the form of the $b_{j,n,i}(a)$ coefficients according to Eq.~\eqref{eq:logexp}.
The approach of Ref.~\cite{Harlander:2009mq,Harlander:2009my} is obtained by ignoring
the second and third terms of the first line of Eq.~\eqref{eq:Ckbest3},
such that the final result is
\begin{align}\label{eq:Ckbest4}
  C^{(k)}\(z,\rhot\)
  &\simeq z^a\sum_{p=0}^{p_{\rm max}} \rhot^p \sum_{i=0}^{i_{\rm max}}(1-z)^i c_{p,i}^{(k)}(a,\ell)
    + \sum_{n=0}^{k-1} A_n^{(k)}(\rhot)\[\frac{\log^nz}{z} - z^a\sum_{i=0}^{i_{\rm max}}(1-z)^i b_{0,n,i}(a) \].
\end{align}
We notice immediately that this form is very similar to Eq.~\eqref{eq:Ckbest2}, with the difference
that the large-$z$ behaviour of the exact small-$z$ contribution is subtracted rather than being damped.
The result is in both cases a small-$z$ contribution which starts at $\Ord\((1-z)^{i_{\rm max}+1}\)$,
and with the same small-$z$ behaviour: it should thus give similar results.
This observation can be considered as an \emph{a posteriori} argument to justify neglecting
the second and third terms of Eq.~\eqref{eq:Ckbest3}.
The sum of these terms isn't necessarily small, and in fact for finite $p_{\rm max}$ it may be sizeable.
The theoretical argument behind neglecting them could be that in the $p_{\rm max}\to\infty$ limit they vanish.
However, the limit is divergent, making such an argument hard to prove.

\paragraph{Method of generalized expansion}

The method of threshold expansion, Eq.~\eqref{eq:Ckbest2},
can be straightforwardly generalized by expanding about a generic $z=z_0$,
\begin{align}\label{eq:Ckbest5}
  C^{(k)}\(z,\rhot\)
  &\simeq z^a\sum_{p=0}^{p_{\rm max}} \rhot^p \sum_{i=0}^{i_{\rm max}} (z_0-z)^i \tilde c_{p,i}^{(k)}(a,z_0)
    + d(z)\sum_{n=0}^{k-1} A_n^{(k)}(\rhot)\frac{\log^nz}{z},
\end{align}
where this time the damping function must be of $\Ord\((z_0-z)^{i_{\rm max}+1}\)$,
in order to avoid spoiling the accuracy of the expansion. Such a function can be
\beq\label{eq:damping5}
d(z) = \(1-\frac{z}{z_0}\)^{i_{\rm max}+1} \theta(z_0-z),
\eeq
where the functional form is such that in $z\to0$ the damping is ineffective,
and the theta function ensures that the small-$z$ contribution remains zero for values of $z>z_0$ where it is forced to vanish.
Eqs.~\eqref{eq:Ckbest5} and \eqref{eq:damping5} reproduce exactly
Eqs.~\eqref{eq:Ckbest2} and \eqref{eq:damping1} for $z_0=1$.\footnote
{To be precise, for $z_0=1$, the coefficients $\tilde c_{p,i}^{(k)}(a,z_0)$ must be replaced with $c_{p,i}^{(k)}(a,\ell)$
which depend on the logarithms $\ell=\log(1-z)$, so the limit $z_0\to1$ isn't smooth.}
For $z_0<1$, the expansion about $z_0$ cannot be valid in a region close to $z=1$, essentially because of the presence
of logarithmic terms diverging in $z=1$ which force the convergence radius to be strictly less than $1-z_0$.
However, this limitation can be simply overcome by patching this result with a purely threshold-expanded one at some $z=z_1$,
with $z_0\leq z_1<1$, to be used for all $z>z_1$.

The advantage of this approach is that the EFT result is used in an extended region of $z$,
while the contribution from the exact small-$z$ terms, which are only known at LL, is relegated to a region of smaller $z$.
The limitation of this approach is that $z_0$ cannot be too small.
Indeed, the EFT approach breaks down for $z\lesssim\rhot/4$, so $z_0$ must be sufficiently larger than this value.
For the physical Higgs and top masses, $\rho_t/4\simeq0.13$.
An interesting value to consider is $z_0=1/2$, for two reasons.
The first is that the EFT expansion parameter, $\rhot/(4z)$, equals $0.26$ in $z=z_0=1/2$, which is just twice as large
as its value at threshold $z=1$, and thus hopefully still sufficiently small for the EFT to be reliable.
The second, more practical, is that the expansions coefficients of the leading EFT contribution ($p=0$)
have been computed for $z_0=1/2$ in the recent work of Ref.~\cite{Mistlberger:2018etf} up to N$^3$LO,
making the implementation of this method rather straightforward.

\paragraph{Conclusion}

The considerations above bring us to the conclusion that the method of threshold expansion, Eq.~\eqref{eq:Ckbest2}, using the damping function Eq.~\eqref{eq:damping1}
and $a=0$ provides the best way of implementing the correct small-$z$ logarithms in a EFT result, such as the NNLO and N$^3$LO ones.
We have implemented this method in the public code \texttt{ggHiggs}, version \texttt{4.0} onwards.
The method of double subtraction, Eq.~\eqref{eq:Ckbest4}, has also been implemented in the code
to test the sensitivity of the results on the method of including small-$z$ contributions
(this method was already used in previous versions of \texttt{ggHiggs} for the NNLO,
with $a=-1$, as prescribed in Ref.~\cite{Harlander:2009mq,Harlander:2009my}).
The method of generalized expansion, Eq.~\eqref{eq:Ckbest5}, with $z_0=1/2$ has been implemented at N$^3$LO
to investigate the possibility of improving the description of the transition region $10^{-2}\lesssim z\lesssim10^{-1}$,
as we will discuss below.
\new{Instead, the method of subtraction, Eq.~\eqref{eq:Ckbest}, due to its implementation difficulties and its arbitrariness, will be discarded.}

The actual numerical implementation of the exact small-$z$ logarithms has to face with the limitation that we know
from resummation only the leading contribution, $A^{(k)}_{n}$ with $n=k-1$, while the coefficients with $n<k-1$ are unknown at NNLO and N$^3$LO.
Here we can follow two possible approaches: we can either include only the known $A^{(k)}_{k-1}$, setting to zero all the other coefficients,
or we can guess their values.
Since the subleading logarithmic contributions are likely more important than the leading one in a region of medium-small $z$,
keeping these coefficients certainly helps.
However, exactly because they may be relevant, their values must be guessed wisely.

To do so, we follow the idea proposed in Ref.~\cite{Ball:2013bra}, namely we include subleading contributions
as predicted by the LL resummation.
To be precise, we use Eqs.~\eqref{eq:Cggexp}, \eqref{eq:Cqgexp} and \eqref{eq:Cqqexp} to construct the $\Ord(\as^2)$ and $\Ord(\as^3)$
contributions to the coefficient functions in the various partonic channels in terms of the coefficients $\tilde{\cal C}_{ij}$.
The anomalous dimensions $\gamma_{0,1,2}$ appearing in those equations are taken to be the expansion terms of the \emph{exact} ``plus'' eigenvalue
of the singlet anomalous dimension matrix, rather than the ones predicted by the resummation, which is appropriate for a fixed-order prediction.
Finally, these expressions are expanded about $N=0$ to identify the resulting $A^{(k)}_n$ coefficients.
This procedure is certainly not fully correct. However, in a NLL (or higher) resummation, these contributions
will be part of the full prediction, which will also contain some additional correction due e.g.\ to the impact factor
at the next perturbative order. The hope is that these corrections be less important that the contributions that we include,
such that the prediction is at least reasonable.

It is clear, however, that such an implementation is not fully satisfactory,
especially at N$^3$LO, where only one out of three parameters is known,
i.e.\ $A^{(3)}_2$ is exact while $A^{(3)}_1$ and $A^{(3)}_0$ are only guessed.
Therefore, it is important to assess, at least qualitatively, the effect of not knowing all the small-$x$ contributions.
To do so, we can consider variations of the unknown parameters.
There are various ways in which these could be done, none of them being particularly justified.
Thus, we consider only a very simple variation, which is obtained by setting to zero the coefficient of $1/z$, $A^{(k)}_0$.
At NNLO, this is the only unknown coefficient, while at N$^3$LO it is the one that governs the
contribution which has the largest impact at medium $z$, and it is thus sufficient to infer
an uncertainty for those values of $z$ relevant for LHC or FCC.
Incidentally, \new{as we shall see, } the resulting uncertainty covers the difference between
the two implementations Eq.~\eqref{eq:Ckbest2} and Eq.~\eqref{eq:Ckbest4},
which is located in the medium\new{/small}-$z$ region, as well as the effect of changing $a$ from 0 to 1 \new{or $-1$. }
Thus, we can take the uncertainty band obtained setting $A^{(k)}_0=0$ as
a good representative of all the small-$z$ uncertainties at fixed order.

\begin{figure}[t]
  \centering
  \includegraphics[width=0.328\textwidth,page=1]{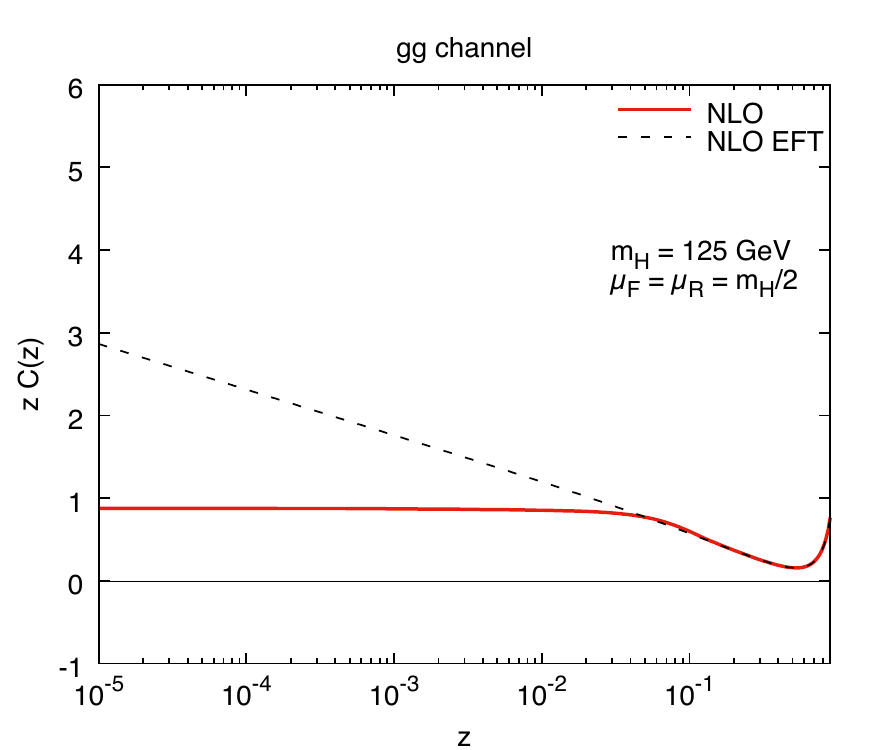}
  \includegraphics[width=0.328\textwidth,page=2]{images/plot_Higgs_partonic_paper}
  \includegraphics[width=0.328\textwidth,page=3]{images/plot_Higgs_partonic_paper}\\
  \includegraphics[width=0.328\textwidth,page=4]{images/plot_Higgs_partonic_paper}
  \includegraphics[width=0.328\textwidth,page=5]{images/plot_Higgs_partonic_paper}
  \includegraphics[width=0.328\textwidth,page=6]{images/plot_Higgs_partonic_paper}\\
  \includegraphics[width=0.328\textwidth,page=7]{images/plot_Higgs_partonic_paper}
  \includegraphics[width=0.328\textwidth,page=8]{images/plot_Higgs_partonic_paper}
  \includegraphics[width=0.328\textwidth,page=9]{images/plot_Higgs_partonic_paper}
  \caption{Partonic coefficient functions for $ggH$ production at fixed order
    (upper plots: NLO; mid plots: NNLO; lower plots: N$^3$LO)
    in solid red and its EFT approximation in dashed black.
    In each row the three plots correspond to the $gg$, $qg$ and $q\bar q$ partonic channels.
    From NNLO onwards there are contributions also from other quark-quark partonic channels,
    which however share the same small-$z$ behaviour and are thus not shown.
    Fixed-order results have an uncertainty band obtained setting $A^{(k)}_0=0$ and symmetrizing the variation.
    \new{At NNLO some variants of the construction based on both Eq.~\eqref{eq:Ckbest2} and Eq.~\eqref{eq:Ckbest4} with $a=-1,0,1$ are shown in solid light blue. }
    At N$^3$LO the alternative implementation Eq.~\eqref{eq:Ckbest5} is also shown in dot-dashed blue.
    The Higgs mass is $\mh=125$~GeV, the top mass $\mt=173$~GeV, and the scales are $\muf=\mur=\mh/2$.}
  \label{fig:ggHpartonicFO}
\end{figure}
In Fig.~\ref{fig:ggHpartonicFO} we show the partonic coefficient functions for the
gluon-gluon, gluon-quark and quark-antiquark partonic channels for factorization and renormalization scales
both equal to half the Higgs mass ($\mh=125$~GeV),
which is nowadays the default central scale adopted by most groups~\cite{deFlorian:2016spz},
and with $\mt=173$~GeV.
In the $gg$ case, only the regular part of the coefficient function is shown, as defined in footnote \ref{foot:reg}.
Results are presented in solid red (NLO in the upper plots, NNLO in the mid plots, N$^3$LO in the lower plots).
At NLO the result is exact~\cite{Spira:1995rr}, while beyond NLO it is constructed according to Eq.~\eqref{eq:Ckbest2}
with $a=0$ and with damping function Eq.~\eqref{eq:damping1}.
Consequently, NNLO and N$^3$LO results are supplemented with an uncertainty band, obtained as described above by setting
$A^{(2)}_0=0$ and $A^{(3)}_0=0$ respectively, and symmetrizing the variation with respect to our central prediction.
For each plot, the leading EFT approximation ($p=0$) is also shown in dashed black.
\new{
At NNLO, we show additional curves which correspond to the two constructions Eq.~\eqref{eq:Ckbest2} and Eq.~\eqref{eq:Ckbest4}
with different values of $a=-1,0,1$.
}
At N$^3$LO, the alternative implementation Eq.~\eqref{eq:Ckbest5} is also shown, together with its own uncertainty band,
in dot-dashed blue.
Note that at N$^3$LO the small-$x$ contributions are different depending on whether the $\MSbar$ or the $Q_0\MSbar$ scheme
is used. Here we decide to use the $Q_0\MSbar$ scheme also at fixed order, to match the scheme adopted
in the resummed results that we will consider in the next subsection.

Several comments are in order.
First, it is apparent that the EFT approximation has the wrong small-$z$ behaviour,
as it exhibits double logarithmic enhancement (at leading power) rather than the correct single logarithmic enhancement.
Indeed, as discussed before, the EFT is expected to fail for $z\lesssim\rhot/4\simeq 0.13$:
this is apparent from the plots, where the red and black curves behave differently for values of $z$
smaller than about $\rhot/4$.
An exception is the $q\bar q$ channel at NLO, where the contribution from the produced $s$-channel gluon
is resonant at the $t\bar t$ threshold in $z=\rhot/4$, producing the peak which is clearly not present in the EFT approximation.
In this case, the agreement between the EFT and exact result is restricted to a region of larger $z$.
This effect is expected to be diluted at higher orders, due to the richer dynamics;
nevertheless it also suggests that it is in general dangerous to trust the EFT result in the vicinity of
$z=\rhot/4$.

The last comment is relevant when analysing the alternative implementation of the N$^3$LO result
based on Eqs.~\eqref{eq:Ckbest5}, \eqref{eq:damping5} with $z_0=1/2$ and $a=0$, blue curve (and band) in the lower plots.
Indeed, as expected, this construction agrees with the EFT result down to lower values of $z\gtrsim0.05$,
thus also reducing the impact of the uncertainty from subleading logarithms in the medium-$z$ region.
However, the considerations above suggest that $z\sim0.05$ is dangerously outside the region of reliability
of the EFT (which is roughly speaking $z>0.2$),
so the gain in precision (smaller uncertainty) of this construction is compensated by a loss in accuracy
(the unknown exact result may lie outside the estimated uncertainty).
This suggests to discard the construction based on Eq.~\eqref{eq:Ckbest5},
and use the safer construction based on Eq.~\eqref{eq:Ckbest2}.

\new{
To study the differences of the other possible alternative constructions proposed earlier in this section,
we have shown in the NNLO plots some curves corresponding to variations of the $a$ parameter
in our default approach Eq.~\eqref{eq:Ckbest2}, and the variant approach Eq.~\eqref{eq:Ckbest4},
again with different values of the $a$ paramenter.
When $a=-1$ (which we consider the lowest acceptable value, even though we favour larger values),
in both approaches the soft expansion produces terms which behave as $1/z$,
and thus differ by a constant amount to our default result in Fig.~\ref{fig:ggHpartonicFO} at small $z$.
This is exactly the form of the subleading contributions used for our estimate of the uncertainty band.
For $a\geq0$, the difference is located in a region of medium $z$, approximately between $z\sim10^{-2}$ and $z\sim10^{-1}$.
Larger values of $a>1$ (not shown in the plots) do not give any visible difference with respect to the results with $a=1$.
Albeit non negligible, these variations are nicely covered by our uncertainty band, as we anticipated.
}

Finally,
at NNLO we observe a reduction of the uncertainty band when going from $gg$ to $qg$ and to the purely quark initiated channel.
This reflects a relatively less important contribution of the small-$z$ logarithms in quark channels.
At N$^3$LO the pattern is the same, but the uncertainty bands are bigger, as appropriate due to the fact that
the fraction of known small-$z$ terms at this order is smaller.
We stress that, in general, the displayed uncertainty is likely an overestimate of the actual uncertainty,
since the coefficient $A^{(k)}_0$ is brutally set to zero rather than varied in a reasonable range.
Thus, the uncertainty band will be useful only to visualize the potential impact of subleading
logarithmic contributions and to motivate further work towards their computation,
rather than for computing an actual uncertainty on the cross section.

\subsection{Impact of high-energy resummation at parton level}
\label{sec:ggHpartonic}

Having described how the exact small-$z$ behaviour is included in fixed-order computations performed
within the large top-mass EFT framework, we now investigate the effect of supplementing the
fixed-order computation with the all-order resummation of small-$z$ logarithms.
At parton level, this is implemented by adding to the fixed-order coefficient functions
the resummed contributions $\Delta_kC_{ij}$ defined in Sect.~\ref{sec:CFexpansion}.
In this section we study the impact of resummation on partonic coefficient functions,
while the effect on the physical cross section will be discussed in the next section.

\begin{figure}[t]
  \centering
  \includegraphics[width=0.328\textwidth,page=1]{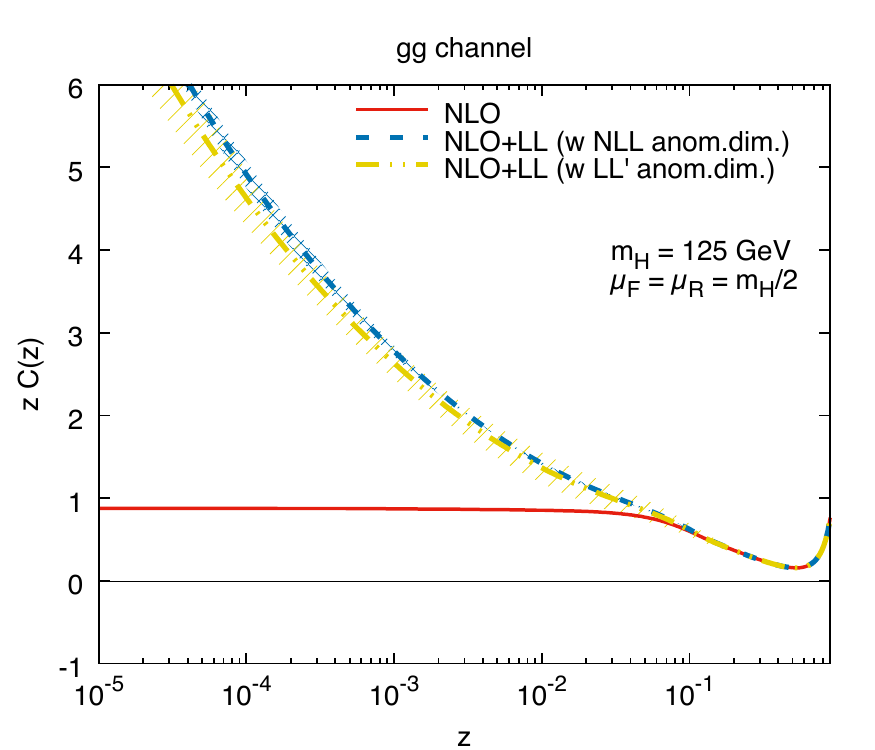}
  \includegraphics[width=0.328\textwidth,page=2]{images/plot_Higgs_partonic_paper_res}
  \includegraphics[width=0.328\textwidth,page=3]{images/plot_Higgs_partonic_paper_res}\\
  \includegraphics[width=0.328\textwidth,page=4]{images/plot_Higgs_partonic_paper_res}
  \includegraphics[width=0.328\textwidth,page=5]{images/plot_Higgs_partonic_paper_res}
  \includegraphics[width=0.328\textwidth,page=6]{images/plot_Higgs_partonic_paper_res}\\
  \includegraphics[width=0.328\textwidth,page=7]{images/plot_Higgs_partonic_paper_res}
  \includegraphics[width=0.328\textwidth,page=8]{images/plot_Higgs_partonic_paper_res}
  \includegraphics[width=0.328\textwidth,page=9]{images/plot_Higgs_partonic_paper_res}
  \caption{Partonic coefficient functions for $ggH$ production at fixed order
    in solid red and fixed order plus resummation in two different implementations:
    dashed blue is the new implementation that uses the NLL anomalous dimension,
    and dot-dot-dashed yellow is the implementation using the LL$^\prime$ anomalous dimension.
    The format and details are as in Fig.~\ref{fig:ggHpartonicFO}.}
  \label{fig:ggHpartonic}
\end{figure}
In Fig.~\ref{fig:ggHpartonic} we report the partonic coefficient functions in the same format as Fig.~\ref{fig:ggHpartonicFO}.
Results are presented at fixed order in solid red,
and with resummation (in the $Q_0\MSbar$ scheme) in the two implementations:
using the NLL anomalous dimension in dashed blue,
and using the LL$^\prime$ one in dot-dot-dashed yellow
(see discussion in Sect.~\ref{sec:evolU} and in Ref.~\cite{Bonvini:2018xvt}).
The fixed-order results are also supplemented by the band which represents a rough estimate
of the potential impact of unknown subleading logarithmic contributions, as described in the previous subsection.
Similarly, the resummed results are supplemented by an uncertainty band,
obtained varying subleading logarithmic contributions related to running coupling
effects in the resummation procedure, as described in Refs.~\cite{Bonvini:2017ogt,Bonvini:2018xvt}.\footnote
{Specifically, we use the sum in quadrature of two independent variations,
one obtained by letting $r(N,\as)\to\as\beta_0$ in Eq.~\eqref{eq:gammaasapprox},
and the other obtained changing the way $\gamma_+$ resums running coupling subleading contributions.}

At NLO and NNLO the two implementations of the resummation give qualitatively similar results,
deviating from the fixed order for $z\lesssim10^{-1}$ at NLO and $z\lesssim10^{-2}\div10^{-3}$ at NNLO.
The growth of the resummed contribution is slightly stronger when the NLL anomalous dimension is used.
The uncertainty band of the LL$'$ variant is slightly larger than the one of the NLL variant,
and covers the latter result everywhere, making them fully compatible.
At NNLO, we notice that the resummed result lies within the fixed-order uncertainty band
for $z\gtrsim10^{-4}\div10^{-3}$, which is the region most important for phenomenology.
If the bands represented faithfully the uncertainty from unknown subleading logarithms at fixed order,
then the effect of resummation in the partonic coefficient functions would be
irrelevant compared to such uncertainty.

At N$^3$LO the general pattern is similar, with some important differences.
The resummed contribution, computed using the NLL anomalous dimension, is a small correction
which lies entirely within the fixed-order uncertainty band for the whole $z$ range shown.
However, this time the behaviour of the resummed result with LL$^\prime$ anomalous dimension
is rather different. In general, the effect is larger than the corresponding one with NLL anomalous dimension,
and no longer fully compatible with it, even though the uncertainty band is also increased.
Moreover, in the $gg$ and $qg$ channels, there is a sizeable contribution of the
resummation in a region of medium-large $z$, $10^{-2}\lesssim z\lesssim0.2$.
This is entirely due to the $\Ord(\as^3)$ expansion of the LL$^\prime$ anomalous dimension $\gamma_2^{\rm LL'}$,
Eq.~\eqref{eq:gamma2LLp}, \new{as explained in Sect.~\ref{sec:orders}, }
and is indeed absent in the quark-quark channel which does not depend on it, see Eq.~\eqref{eq:Cqqexp}.
This large contribution is in a region of $z$ which cannot be considered
to be dominated by small-$z$ logarithms, and therefore has to be interpreted as a spurious effect.
Indeed, the all-order resummed results with NLL and LL$^\prime$ anomalous dimensions agreed in that $z$ region when matched to NLO and NNLO,
so there is no physical underlying reason for which they should give such different results when matched to one order higher.

Our interpretation of the origin of this spurious behaviour is the fact that while the LL$^\prime$ anomalous
dimension makes perfect sense to all orders, its $\as$ expansion may be unstable order by order,
perhaps due to its hybrid nature, \new{and to the fact discussed in Sect.~\ref{sec:orders} that
none of the non-vanishing contributions at $\Ord(\as^2)$ and $\Ord(\as^3)$ is exact. }
This is not the case for the NLL anomalous dimension, which has a well behaved $\as$ expansion,
\new{with the leading non-vanishing singularity correctly predicted at each order. }
This conclusion is in agreement with the analysis of Ref.~\cite{Bonvini:2018xvt},
and represents another motivation for favouring the use of the NLL anomalous dimension
in place of the LL$^\prime$ one in the computation of resummed coefficient functions,
in particular when these are matched to N$^3$LO or to a higher order.
Nevertheless, we must bear in mind that the resummed result based on the LL$'$ anomalous dimension
differs from the NLL one by \new{formally } subleading contributions.
Therefore, the difference between the two formulations probes unknown subleading logarithms.
We see that this difference is similar (slightly more conservative) than the uncertainty band
on the NLL-based resummed result when matched to NLO or NNLO,
and could thus be used as an alternative way of estimating subleading logarithmic uncertainty.
When matched to N$^3$LO, this difference is rather larger than the simple blue band,
especially in the medium-large $z$ region, and using it as a subleading logarithmic uncertainty
would be rather conservative.
However, given that we do not really know how large these subleading logarithms may actually be,
we suggest to use this difference as a measure of such uncertainty.
As we will see in the next section the resulting uncertainty at the physical cross section level is very reasonable.

\begin{figure}[t]
  \centering
  \includegraphics[width=0.49\textwidth,page=1]{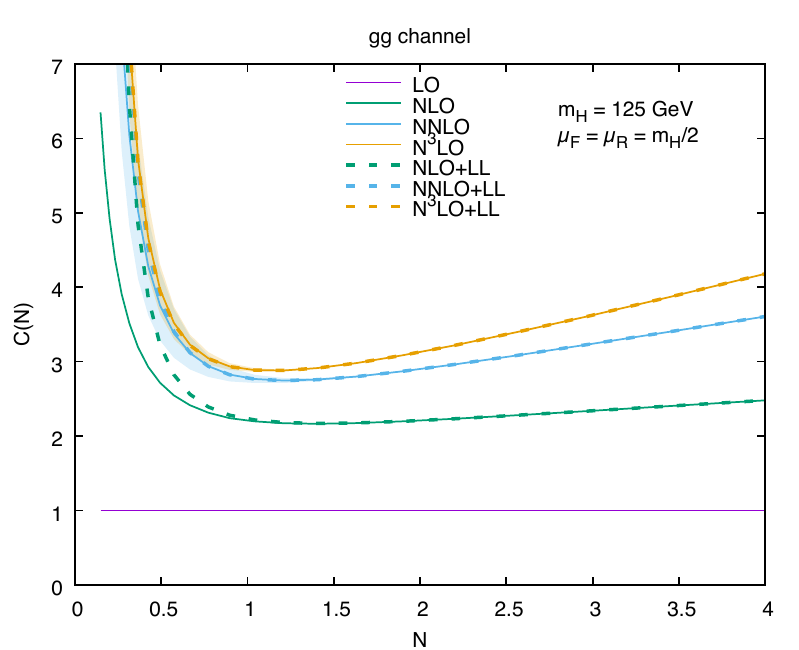}
  \includegraphics[width=0.49\textwidth,page=2]{images/plot_Higgs_partonic_Nspace}
  \caption{Partonic $gg$-channel coefficient functions for $ggH$ production in $N$ space,
    at LO (solid purple), NLO (solid green), NNLO (solid blue) and N$^3$LO (solid orange)
    and with resummation at NLO+LL (dashed green), NNLO+LL (dashed blue) and N$^3$LO+LL (dashed orange).
    The left plots shows the actual coefficient functions, while the right plot shows their ratio
    to the highest-order result, N$^3$LO+LL.
    Fixed-order results are supplemented with the uncertainty band obtained setting $A^{(k)}_0=0$ and symmetrizing the variation;
    the band at N$^3$LO is just the one from the $\Ord(\as^3)$ contribution
    and does not contain the contribution from the previous order.
    Similarly, the uncertainty bands on the resummed contributions (right plot only) are
    estimated as the difference between the NLL and LL$'$ variants of the resummation.
    The scales are $\muf=\mur=\mh/2$.}
  \label{fig:ggHpartonicNspace}
\end{figure}
Another powerful way of visualizing the effect of resummation at parton level
is through the Mellin transform of the coefficient functions.
In Fig.~\ref{fig:ggHpartonicNspace} we show the dominant one, $C_{gg}$, as a function of the Mellin variable $N$
for positive real $N$.
This time we include the full coefficient function, and not just the regular part,
since the Mellin transform of a distribution is an ordinary function.
In fact, the distributional part of the coefficient function is responsible for the growth of the
coefficient function at large $N$~\cite{Bonvini:2012sh}.
Moreover, it is known~\cite{Bonvini:2012an,Bonvini:2012sh,Bonvini:2014qga}
that a saddle point evaluation of the Mellin inversion integral
defining the full cross section (i.e.\ including both the coefficient functions and the PDFs)
provides an excellent approximation to the exact result, thus showing that the bulk
of the contribution of the coefficient function to the cross section is encoded in its value at the saddle point $N=N_{\rm saddle}$.
From Ref.~\cite{Bonvini:2012an} we know that the saddle point for SM Higgs production varies from\footnote
{Note that in the mentioned references a different, more standard definition of the Mellin transform
is used where the leading high-energy singularity is in $N=1$. In this work we use a different definition,
common in high-energy resummation literature, where the leading singularity is in $N=0$.
Thus the values of $N$ read from those references must be lowered by a unity.}
$N_{\rm saddle}\simeq1.1$ for LHC at $\sqrt{s}=7$~TeV to
$N_{\rm saddle}\simeq0.9$ for LHC at $\sqrt{s}=14$~TeV and to
$N_{\rm saddle}\simeq0.7$ for FCC at $\sqrt{s}=100$~TeV.
Thus, the region of interest for phenomenology in a vast range of hadron-hadron collider energies is all
located in a small range of $N$ close to $N=1$.

In the left plot of Fig.~\ref{fig:ggHpartonicNspace} the full LO, NLO, NNLO and N$^3$LO coefficient functions are shown,
together with the resummed NLO+LL, NNLO+LL and N$^3$LO+LL counterparts.
Since the plot becomes busy in the small-$N$ region, we also plot in the right panel
the ratio of each curve to the highest order curve, N$^3$LO+LL.
We see that the resummed results depart from the fixed order for $N<1$,
and they all diverge at the same $N=N_{\rm pole}>0$, which is determined by the resummation.
Thus, they all grow stronger than each fixed order, which instead are singular in $N=0$.
Interestingly, the N$^3$LO+LL curve is very close to the N$^3$LO curve even at rather small $N\gtrsim0.2$,
which is in line with the behaviour found in the $z$-space plots, and shows that the effect of small-$z$ resummation
on the N$^3$LO coefficient function is expected to be negligible,
since the saddle point is in a region where N$^3$LO and N$^3$LO+LL are almost identical.
In particular, the effect of subleading logarithms at fixed order, estimated by the coloured filled bands in the plots,
is likely more significant than the effect of all-order resummation, both at NNLO and N$^3$LO.
The fixed-order uncertainty bands also appear to be larger than the uncertainty on the resummed contributions,
estimated as the difference between the NLL and LL$'$ variants, and shown with a pattern.
While we may hope, as already discussed, that these bands be over conservative,
it seems important to take this observation as a strong motivation to work towards improving the knowledge
of the small-$z$ behaviour of the Higgs partonic coefficient functions.

\subsection{Impact of high-energy resummation on the cross section}
\label{sec:ggHxs}

We now move to the physical cross section.
It is defined as the convolution of the partonic coefficient functions with the PDFs,
according to Eq.~\eqref{eq:sigmaColl} which in momentum space reads\footnote
{Note that in the case of the Higgs cross section $\sigma_0$ is independent of $N$ in Mellin space,
and thus it factors out also in the Mellin convolution in momentum space.
Additionally, the sum extends over all quark flavours and not just the singlet combination.}
\begin{align}\label{eq:sigmaCollzspace}
\sigma(N,Q^2) &= \sigma_0(Q^2) \sum_{i,j=g,q} \int_\tau^1 \frac{dz}z\,
C_{ij}\(z,\as(\mur^2),\frac{\muf^2}{Q^2},\frac{\mur^2}{Q^2}\)\, \Lum_{ij}\(\frac{\tau}{z},\muf^2\),\\
\Lum_{ij}(x,\muf^2) &= \int_x^1 \frac{dy}{y} \, f_i(y,\muf^2)\, f_j\(\frac{x}{y},\muf^2\), \label{eq:Lum}
\end{align}
with $\tau=\mh^2/s$ and $s$ che collider center-of-mass energy,
and we have restored the dependence on the renormalization scale $\mur$.
Since high-energy resummation affects PDF evolution, and PDFs at small-$x$ are mostly determined
by HERA data at low $Q^2$ which are thus very sensitive to resummation effects and very ``far''
from the Higgs scale,
it is crucial to use PDFs which have been determined and evolved using resummed theory
when computing physical predictions which include high-energy resummation.

Recently, such PDFs have been determined in the context of the NNPDF methodology to PDF fitting~\cite{Ball:2017otu}.
Soon after, the xFitter collaboration also performed an analogous determination~\cite{Abdolmaleki:2018jln},
whose findings are in agreement with those of the NNPDF study.
In both cases, PDF sets have been fitted using fixed-order theory (NLO or NNLO\footnote
{The xFitter study~\cite{Abdolmaleki:2018jln} only considered NNLO theory,
  since the effects of small-$x$ resummaiton are more marked at that order.})
supplemented by high-energy resummation at NLL in the $Q_0\MSbar$ scheme provided by the \texttt{HELL} code, version \texttt{2.0}.
To be precise, resummation in DGLAP evolution is really NLL, while resummation 
in DIS coefficient functions is just formally NLL, since the LL contribution vanishes.
In this case, we would refer to the accuracy of resummation in DIS as \emph{absolute} NLL but \emph{relative} LL
(for this notation, see Ref.~\cite{Bonvini:2017ogt}).
In this respect, Higgs resummation, which is \emph{relative} LL,
is consistent with the PDF sets of Refs.~\cite{Ball:2017otu,Abdolmaleki:2018jln}.

In fact, since the Higgs cross section is known at fixed order up to N$^3$LO,
a consistent computation would require the use of PDFs obtained with N$^3$LO theory,
supplemented by resummation when computing resummed cross sections.
However, this would require four-loop DGLAP splitting functions,
which are not known yet, even though recently there has been some impressive progress
towards their computation~\cite{Davies:2016jie,Moch:2017uml,Vogt:2018ytw}.
Therefore, for the time being we can only rely on NNLO (or NNLO+NLL) PDFs.

We will focus on the PDFs of Ref.~\cite{Ball:2017otu}, which are publicly available.
In that work, various families of PDF sets have been obtained by using different datasets.
The mainstream family is based on a global dataset, which includes on top of DIS data a large amount of
``hadronic'' data (mostly Drell-Yan, jet and $t\bar t$ production),
selected in a region where resummation effects in the coefficient functions are expected to be negligible,
since for these observables resummation is not yet available in \texttt{HELL}.
Another family is then obtained by including only the DIS datasets in the fit,
such that resummation is consistently included for all datapoints.
Three variants of these DIS-only fits have been created by enlarging the dataset to include
pseudo-data from possible future DIS experiments, namely the Large Hadron-electron Collider (LHeC),
the Future Circular electron-hadron Collider (FCC-eh), and both.

For each family, four fits have been performed, with NLO, NLO+NLL, NNLO and NNLO+NLL theory
(except for the LHeC and FCC families where only NNLO and NNLO+NLL is available).
In all cases, the resummation makes use of the LL$^\prime$ anomalous dimension,
which, as suggested in Ref.~\cite{Bonvini:2018xvt} and confirmed in this work, is not the best choice.
The new version of \texttt{HELL} released with this work, \texttt{3.0},
uses the NLL anomalous dimension rather than the LL$^\prime$ one as default,
so future fits including high-energy resummation will be performed with this new setup.
So far, only a single PDF set with resummation at NNLO+NLL has been determined using the NLL anomalous dimension,
which has been used in Ref.~\cite{Ball:2017otu} to investigate the effect of subleading logarithms.
However, this set is based on the DIS-only dataset, and as such it suffers from larger uncertainties
and it is then not suitable for phenomenology.
Nevertheless, its existence will be helpful to investigate the effect of computing consistently
the Higgs cross section with our favourite choice of NLL anomalous dimension.

We have to warn the Reader that the \texttt{HELL 2.0} version of the code~\cite{Bonvini:2017ogt}
used for the aforementioned PDF fits was based on an incorrect resummation formula,
which produced spurious NLL contributions to the $P_{gg}$ splitting function beyond $\Ord(\as^3)$,
and affected other splitting functions and coefficient functions beyond their logarithmic accuracy.
The issue has been corrected in \texttt{HELL 3.0}~\cite{Bonvini:2018xvt}.
The effect of the correction at the level of splitting functions and coefficient functions
appears to be reasonably small
~\cite{Bonvini:2018xvt},
especially in the kinematic region of HERA, so we expect that the resulting PDFs are not
severely affected by the issue.
We stress however that the difference between the LL$'$ and NLL formulations of the resummation matched to NLO or NNLO
is significantly reduced after the correction: therefore, the non-negligible difference in the PDFs~\cite{Ball:2017otu}
obtained with these two formulations using the previous version of the code
will likely be reduced significantly in future PDF fits based on \texttt{HELL 3.0}.

\new{
The effect of including resummation in the theory used for PDF determination is
on the one hand an improvement of the quality of the description of the data,
and on the other hand a rather different gluon and quark-singlet PDFs at small $x$.
Such effect is much larger when resummation is added on top of the NNLO than on the NLO.
The resulting gluon and quark-singlet PDFs at NNLO+NLL are harder at small-$x$ than their NNLO counterparts.
The shape of the resummed PDFs is very similar in both Ref.~\cite{Ball:2017otu} and Ref.~\cite{Abdolmaleki:2018jln},
despite some important differences in the fitting methodology, the dataset and the treatment of the charm PDF.
It is important to stress that the data constraining the PDFs at small $x$ are mostly inclusive HERA data~\cite{Abramowicz:2015mha}
which lie at a small energy scale $Q^2$. The effect of small-$x$ resummation in the fit of PDFs
is thus induced by the modified description of the DIS structure functions at low $x$ and $Q^2$,
which in turn determines a different gluon and quark-singlet PDFs at low $Q^2$,
which is then evolved to higher scales through DGLAP evolution (using resummed splitting functions).
Therefore, the effect of resummation on PDFs at small $x$ at the Higgs scale is somewhat indirect
(this is true also for fixed-order PDFs at small $x$), though not less reliable.
However, it would prove very useful to include in future additional data at small $x$ and larger $Q^2$,
e.g.\ from forward Drell-Yan at LHCb, to further constrain the small-$x$ PDFs at a scale closer to the Higgs scale.
The resummation of such a process in \texttt{HELL} is work in progress.
}

In the rest of this section we will proceed as follows.
First, we take the global PDF sets of Ref.~\cite{Ball:2017otu}
and compute predictions for the resummed Higgs cross section.
Then, we will use the DIS-only PDFs to study the impact of subleading terms,
both at the level of PDFs and of the coefficient functions.
Additionally, in the context of the DIS-only sets we will investigate the reduction of the PDF uncertainty
on the Higgs cross section that could be achieved with future DIS experiments.

\begin{figure}[t]
  \centering
  \includegraphics[width=0.49\textwidth,page=2]{images/ggH_Kfactor_summary2}
  \includegraphics[width=0.49\textwidth,page=1]{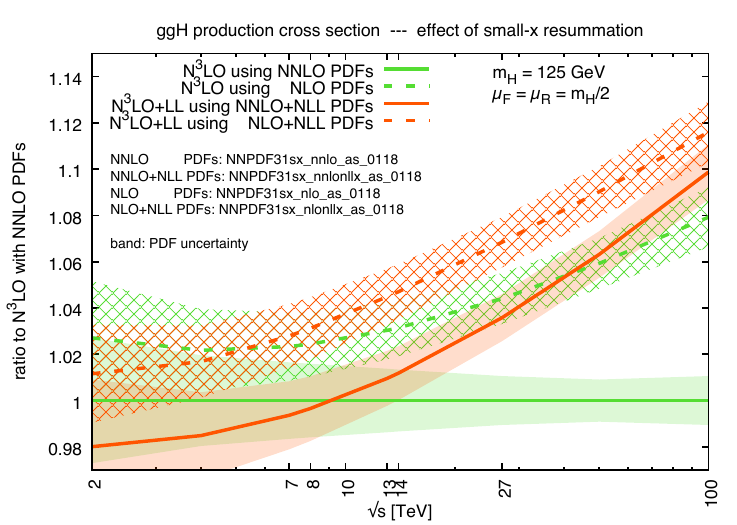}
  \caption{Ratio of the N$^3$LO Higgs cross section with and without resummation to the N$^3$LO fixed-order cross section,
    as a function of the collider center-of-mass energy.
    The PDFs used are from the global dataset of Ref.~\cite{Ball:2017otu}.}
  \label{fig:ggHkfactor}
\end{figure}
Let us start with the PDFs based on the global dataset.
We consider $\mh=125$~GeV (physical Higgs) and $\mt=173$~GeV, and compute the cross section
as a function of the collider center-of-mass energy $\sqrt{s}$.
We set the scales to $\mur=\muf=\mh/2$, which is our default central choice.
In Fig.~\ref{fig:ggHkfactor} (left plot) we show the cross section at N$^3$LO and N$^3$LO+LL
for a range of collider energies which spans from a Tevatron-like\footnote
{We are assuming that the collider is a proton-proton collider, so this prediction is not really a Tevatron prediction.
However, the difference between proton and antiproton PDFs is limited to non-singlet PDFs,
which give a negligible contribution to the Higgs cross section.}
energy of $\sqrt{s}=2$~TeV to a FCC-hh energy of $\sqrt{s}=100$~TeV.
Since the cross section changes significantly over this large range of energies,
we present the results as ratios ($K$-factors) with respect to the fixed-order N$^3$LO prediction.
For the fixed order (green) and resummed (red) predictions we use
the NNLO and NNLO+NLL global PDF sets of Ref.~\cite{Ball:2017otu}, respectively.
The uncertainty band shown represents the PDF uncertainty only.
We see that the effect of resummation is small and compatible within the PDF uncertainty
for small collider energies, up to the current LHC energy of $\sqrt{s}=13$~TeV.
From this value onward the net effect of the resummation is a significant increase of the cross section
with respect to the fixed-order prediction, reaching up to $+10\%$ for FCC at $\sqrt{s}=100$~TeV.

This huge effect may seem surprising, and thus deserves a careful investigation.
First of all, we note that basically the whole effect comes from the use of resummed PDFs,
while the effect of the resummation in the coefficient function is almost negligible.
Indeed, in the same plot there is an additional curve (dashed blue) obtained by computing the fixed-order N$^3$LO cross section
with the resummed PDFs: this curve, which differs from the red one only by the resummed contributions to the
coefficient function, is basically identical to it, except for a tiny deviation visible only at large collider energies
grater than $\sqrt{s}\sim30$~TeV.
These observations naturally raise the following questions.
Why is the effect of high-energy resummation in the PDFs and in the partonic coefficient functions so unbalanced?
Specifically, why is the effect of resummation in the PDFs so large?
And why is the effect of resummation in the partonic coefficient functions so small?
We now answer these three questions in turn.

\begin{figure}[t]
  \centering
  \includegraphics[width=0.328\textwidth,page=1]{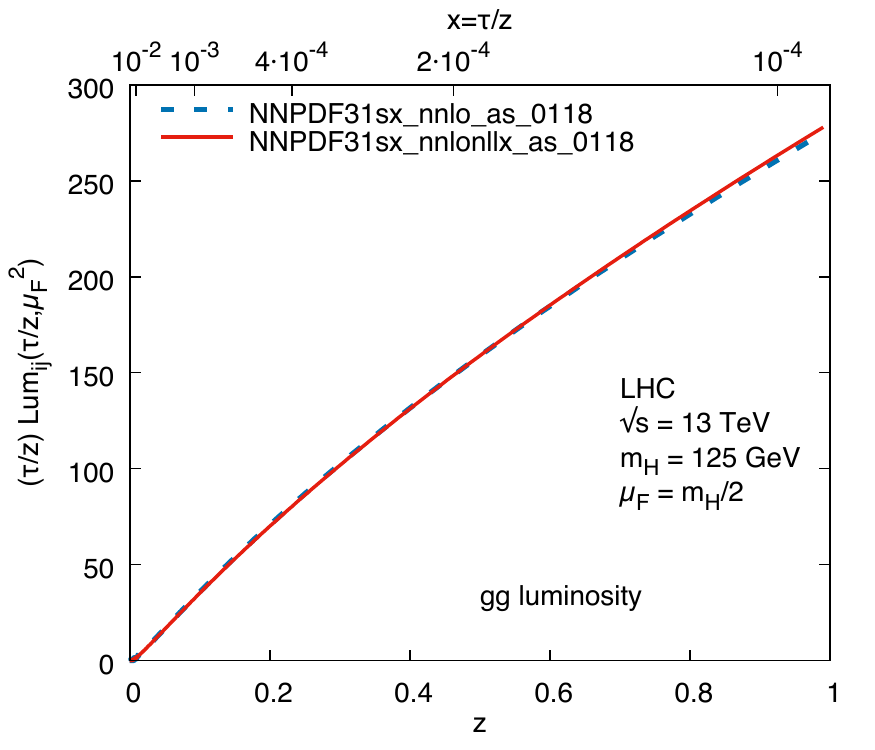}
  \includegraphics[width=0.328\textwidth,page=2]{images/plot_Higgs_lumi-plot}
  \includegraphics[width=0.328\textwidth,page=3]{images/plot_Higgs_lumi-plot}
  \includegraphics[width=0.328\textwidth,page=4]{images/plot_Higgs_lumi-plot}
  \includegraphics[width=0.328\textwidth,page=5]{images/plot_Higgs_lumi-plot}
  \includegraphics[width=0.328\textwidth,page=6]{images/plot_Higgs_lumi-plot}
  \caption{The luminosities $x\Lum_{ij}(x,\muf^2)$ for $x=\tau/z$ as a function of $z$,
    for $ij=gg$ (left plot), $ij=qg$ (middle plot) and $ij=q\bar q$ (right plot),
    for Higgs production at LHC at $\sqrt s=13$~TeV (top row) and FCC-hh at $\sqrt s=100$~TeV (bottom row).
    Values of $x=\tau/z$ are shown in the upper horizontal axis.
    The factorization scale corresponds to the central choice $\muf=\mh/2$.
    The PDFs used are from the global dataset of Ref.~\cite{Ball:2017otu}.}
  \label{fig:lumi}
\end{figure}
The unbalance between the effect of resummation in PDFs and partonic coefficient functions
is a characteristic feature of the observable under consideration being an inclusive cross section,
and is due to the form of the convolution defining the cross section, Eq.~\eqref{eq:sigmaCollzspace}.
In particular, given that in the convolution when the coefficient functions are computed in $z$ the PDF luminosities
are computed in $x=\tau/z$, in the integration small-$z$ coefficient functions multiply large-$x$ PDFs and vice versa.
To illustrate why this generates an unbalance, we show in Fig.~\ref{fig:lumi} the luminosities $x\Lum_{ij}(x,\muf^2)$
for $x=\tau/z$, as a function of the integration variable $z$,
for $ij=gg$ (left plot), $ij=qg$ (middle plot) and $ij=q\bar q$ (right plot).
These functions are the weights to the coefficient functions
in the integral Eq.~\eqref{eq:sigmaCollzspace} defining the cross section.
Since such functions depend on $\tau=\mh^2/s$, we show both the case for current LHC ($\sqrt s=13$~TeV, first line)
and FCC-hh ($\sqrt s=100$~TeV, second line).
It is clear that when the integration variable $z$ is small,
and thus the small-$z$ logarithms are enhanced in the coefficient functions,
the parton luminosities (and thus the PDFs) are computed at large values of their argument $x=\tau/z$
(reported in the upper axis),
where the PDFs vanish, giving a suppressed contribution to the integral.
Therefore, the region where small-$z$ resummation has an effect in the coefficient functions
(roughly $z\lesssim10^{-2}$, from Fig.~\ref{fig:ggHpartonic}), gives a tiny contribution to the convolution integral,
i.e.\ to the cross section.
On the contrary, the large-$z$ (threshold) region (roughly speaking, the region $z\gtrsim0.1$)
is enhanced in the integrand by the larger value of the luminosities and dominates the integral.\footnote
{This enhancement of the large-$z$ portion of the integrand due to the PDF luminosities is
  a well known effect~\cite{Becher:2007ty,Bonvini:2012an,deFlorian:2014vta,Contino:2016spe},
  and it is the reason for which threshold (large-$z$) resummation is important for this process.}
In this region, the resummed coefficient functions reduce to their fixed-order limit (Fig.~\ref{fig:ggHpartonic})
and are thus insensitive to small-$z$ resummation,
but the PDFs are computed at smaller values of their argument, and are thus potentially sensitive to small-$x$ logarithmic enhancement.
Since this region is enhanced by the larger values of the luminosities, the effect of small-$x$ resummation in PDFs,
if present, is enhanced with respect to the effect of small-$z$ resummation in coefficient functions.
Indeed, in the plots the luminosities are computed using both the NNLO (dashed blue) and the NNLO+NLL (solid red)
sets of PDFs of Ref.~\cite{Ball:2017otu}, and it is apparent that in the FCC case, which probes smaller values of $x=\tau/z$,
all the luminosities are very different at large $z$, giving the aforementioned $10\%$ effect on the cross section.
In the LHC case, the discrepancy between the two PDF sets is much less marked,
but still sufficient to give the $1\%$ effect observed in Fig.~\ref{fig:ggHkfactor}.

Regarding the second question,
we argue that the origin of this huge difference between the predictions obtained with either the NNLO or the NNLO+NLL PDFs
is due to the former being unreliable at small $x$, due to a perturbative instability in the splitting functions
and DIS coefficient functions at NNLO, in turn due to the unresummed small-$x$ logarithms.
Indeed, in Ref.~\cite{Ball:2017otu} it was observed that the behaviour of the NNLO gluon PDF at small $x$ is rather
different from that of the NLO PDF; the latter, in turn, is quite similar to both the NLO+NLL and NNLO+NLL resummed gluon PDFs.
Namely, the perturbative progression of the PDFs is perturbatively stable at small $x$ when resummation is included,
but unstable when resummation is not included, the instability starting to appear at NNLO.
To understand how much of this PDF behaviour is reflected on the Higgs cross section,
we show in Fig.~\ref{fig:ggHkfactor} (right plot) the fixed-order and resummed cross sections using
NLO PDFs (dashed green), NNLO PDFs (solid green), NLO+NLL PDFs (dashed red) and NNLO+NLL PDFs (solid red).
We observe indeed that at high collider energies
(which probe smaller $x$ and are thus more sensitive to small-$x$ logarithms and their resummation)
all curves except the one with NNLO PDFs are grouped together,
indicating that the small-$x$ instability of the NNLO is really the culprit of the huge difference between
fixed-order and resummed results at high collider energies.
Indeed, the resummed result with NNLO+NLL PDFs is a reasonably small correction to the results obtained
with either NLO or NLO+NLL PDFs at high energies.
We conclude that the effect of small-$x$ resummation on the Higgs cross section is per se not surprisingly large;
however, using NNLO PDFs gives rise to unreliable results at high energies, due to the instability at small-$x$,
which is not even covered by the PDF uncertainty.
This effect is expected to be even more marked with N$^3$LO PDFs, since N$^3$LO splitting functions
suffer from stronger instabilities, as demonstrated in Ref.~\cite{Bonvini:2018xvt}.
Thus, contrary to the common lore, using N$^3$LO PDFs for a N$^3$LO cross section
such as the Higgs cross section would produce results which are even less reliable than those with lower order PDFs.
Therefore, at high energies precise and reliable predictions can only be based on small-$x$ resummed PDF sets.

\new{
We also observe that at small collider energies using resummed NNLO+NLL PDFs
gives a reduction of the cross section, which seems to approach a constant value of about $-2\%$.
Here the PDF uncertainties are large, and with respect to them this effect is not significant.
Moreover, the Higgs cross section at these energies is so small to be not phenomenologically relevant.
Even so, it is interesting to explain the origin of this effect.
This reduction of the cross section originates from a depletion of the gluon PDF for $10^{-2}\lesssim x\lesssim10^{-1}$
when resummation is included, see e.g.\ Fig.~4.8 of Ref.~\cite{Ball:2017otu}.
While this effect is not genuinely a small-$x$ effect, its origin is indirectly due the inclusion
of small-$x$ resummation through the contraint imposed in the PDF fits by the momentum sum rule:
the smaller gluon at medium/large $x$ compensates the larger gluon at small $x$.
It is then important to keep in mind that even though small-$x$ resummation has 
its largest effects at small $x$, the changes in the theoretical ingredients of PDF fits
also induce (smaller) effects at medium and large $x$, which in turn may lead to visible effects
on some observables not directly sensitive to the small-$x$ region.
}

Moving to the third question,
we now return to the observation that the resummation in the coefficient function has a tiny effect.
This fact is partly due to the fact that we are adding resummation in the $Q_0\MSbar$ scheme on top of the already rather precise
N$^3$LO prediction, and is in perfect agreement with the parton-level behaviour observed in Sect.~\ref{sec:ggHpartonic},
together with the observation that the Higgs cross section is threshold dominated, as clear from Fig.~\ref{fig:lumi}.
However, the size of small-$z$ contributions to the coefficient functions may be different when treating differently
subleading contributions, or in different factorization schemes.
Indeed, we have noted in Sect.~\ref{sec:ggHpartonic} that the partonic behaviour is rather different when N$^3$LO is supplemented
with the resummation computed with the LL$^\prime$ anomalous dimension.
In that case, the resummation has an effect also at larger $z$, and may then survive the luminosity suppression.
While we believe that this effect is spurious, it is interesting
to see how it affects the physical cross section.
This is also interesting because the PDF sets of Ref.~\cite{Ball:2017otu} have been obtained
using resummation based on the LL$^\prime$ anomalous dimension.\footnote
{We recall that the PDF set was obtained with the previous version of \texttt{HELL},
  and therefore the LL$'$ is not really consistent with those PDFs.}
(However, the instability of the LL$^\prime$ anomalous dimension appears when expanded to $\Ord(\as^3)$,
which is not the case for the NNLO+NLL resummation used in the PDFs,
for which the use of the LL$'$ formulation can be considered reliable.)
Thus, in the left plot of Fig.~\ref{fig:ggHkfactor} we also show the LL$^\prime$ version
of the resummed prediction (dot-dashed red).
In this case, the effect is rather large, even for small collider energies where we expect resummation to have no effect:
this is entirely due to the sizeable contribution of the resummation at $z\sim10^{-1}$ (see Fig.~\ref{fig:ggHpartonic}),
and confirms the spurious nature of such effect.
(Interestingly, the effect is positive, i.e.\ it does not compensate in any way the effect of the resummation in the
PDFs, which would be expected if the process were included in the PDF fit.)
However, it also points out that a different treatment of subleading contributions may give sizeably different results,
so the smallness of the effect of resummation in the coefficient functions is also due to the specific choice of using
the NLL anomalous dimension in the resummation.

While we have found strong motivations to discard the resummation based on the LL$^\prime$ anomalous dimension,
we have suggested in Sect.~\ref{sec:ggHpartonic} to use the difference of the resummed predictions
obtained with NLL and LL$^\prime$ anomalous dimension as an uncertainty due to unknown subleading logarithmic contributions.
This choice is certainly conservative if one considers the effect on the coefficient function alone.
However, we shall not forget that subleading logarithmic contributions may have sizeable effects in the PDFs as well,
which are probably not taken into account by the PDF uncertainty (see also discussion in Ref.~\cite{Ball:2017otu}).
Thus, this uncertainty has the role to also account for subleading logarithms in PDFs,
e.g.\ to compensate for the fact that these PDFs have not been obtained using the NLL anomalous dimension.

It would be interesting to quantify how large the uncertainty from subleading logarithmic contributions in the PDFs can be.
One way to do so is to use a PDF set which has been determined using resummation implemented through the NLL anomalous dimension.
\new{
In such a PDF set both the DGLAP evolution and the theory used to describe the DIS data at small $x$
(all of which lie at small $Q^2$) differ by subleading contributions with respect to
the implementation based on the LL$^\prime$ anomalous dimension.
Despite the fact that the difference is subleading, and that when resummation is matched to NNLO
(as in the PDF fit) the difference between the LL$'$ and NLL implementations is small,
the effect on the resulting PDFs may be sizeable, mostly because the relevant DIS data lie at small $Q^2$
where higher order corrections are enhanced by large values of $\as(Q^2)$.
As anticipated, in Ref.~\cite{Ball:2017otu} a single NNLO+NLL fit based on the NLL anomalous dimension\footnote
{However, as already mentioned, also this fit was performed prior to the correction in the resummation code,
where the difference between the LL$'$ and NLL variant was larger than in the bug-fixed version \texttt{HELL 3.0}.}
has been performed.
As one can appreciate from Fig.~4.4 of Ref.~\cite{Ball:2017otu}, the qualitative behaviour of the PDFs
and the significance of the effect of resummation is the same with both choices for the anomalous dimension.
Nevertheless, the effect of subleading contributions gives a quantitatively different result,
as one may expect from the argument above.
This effect is not covered by the PDF uncertainty, and thus it is important to understand how it impacts a physical cross section.
However, this variant of the fit was performed just in the context of the DIS-only dataset.
}
Thus, to investigate the effects of subleading contributions in a consistent manner,
we need to consider the DIS-only fits, which however suffer from larger uncertainties
and are thus not suitable for phenomenological applications.

\begin{figure}[t]
  \centering
  \includegraphics[width=0.49\textwidth,page=3]{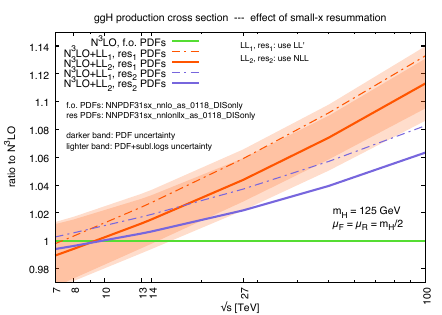}
  \includegraphics[width=0.49\textwidth,page=4]{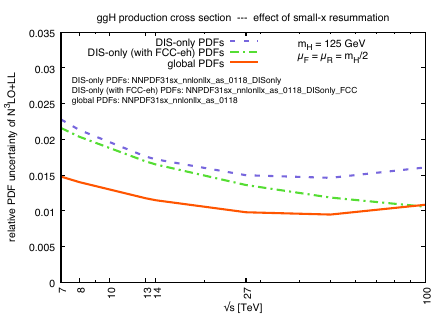}
  \caption{Left plot: ratio of the N$^3$LO+LL Higgs cross section to the N$^3$LO one as a function of the collider energy,
    using various combinations for the implementation of resummation in the coefficient function and in the (DIS-only) PDFs.
    Right plot: relative PDF uncertainty as a function of the collider energy for the resummed cross section obtained
    with the global, DIS-only and DIS-only+FCC-eh PDF sets of Ref.~\cite{Ball:2017otu}.}
  \label{fig:ggHkfactorDIS}
\end{figure}
In Fig.~\ref{fig:ggHkfactorDIS} we show (left plot) the resummed cross section
(normalized to the N$^3$LO one computed with NNLO PDFs) with four different combinations
of choices of subleading contributions:
using consistently the LL$^\prime$ anomalous dimension in both PDFs and coefficient functions (dot-dashed red),
using the LL$^\prime$ anomalous dimension in the PDFs and the NLL one in the coefficient functions (solid red, our default),
using consistently the NLL anomalous dimension in both PDFs and coefficient functions (solid blue),
and using the NLL anomalous dimension in the PDFs and the LL$^\prime$ one in the coefficient functions (dot-dashed blue).
We restrict our attention to the LHC--FCC energy range, and show on our default prediction (solid red)
the PDF uncertainty band (darker red area) and the sum in quadrature of it with the ``subleading logarithmic uncertainty''
as defined above, namely by the difference between solid and dot-dashed red (lighter red area).
The solid blue curve is what we would consider the new default prediction,
as it uses consistently the NLL anomalous dimension, as suggested in Ref.~\cite{Bonvini:2018xvt} and here.
We note that such prediction is smaller than our default one,
reaching ``just'' a $6\%$ increase over the N$^3$LO at FCC,
and suggesting that our current default prediction may overestimate the real effect.
Nevertheless, we see that our full uncertainty band reasonably takes into account
the difference between the two predictions, even though the blue curve lies outside the band for $\sqrt{s}\gtrsim30$~TeV.
However, we need to keep in mind that these PDFs are based on the previous version of \texttt{HELL},
where the difference between LL$'$ and NLL formulations was larger than in the new corrected version, as we commented before.
We may realistically expect that with the new version of the code the PDF sets corresponding
to the two variants of the resummaiton be closer to each other, such that our uncertainty band
successfully covers such effect.
A definitive answer can only be obtained in future, when a (possibly global) PDF fit
will be performed with the new \texttt{HELL 3.0} default, ideally also including the
resummation of hadron-hadron collider observables, most importantly Drell-Yan cross sections,
which can directly constrain small-$x$ PDFs at larger $Q^2$ and then reduce an unavoidable source
of uncertainty coming from the large portion of DGLAP evolution from the low-$Q^2$ HERA region
(where the data which constrain the PDFs at low $x$ lie) to the ElectroWeak scale.
\new{In any case, it appears clear that subleading contributions at small $x$ are important,
and should be taken into account when computing the uncertainty from missing higher orders in PDF determination.}

In the context of DIS-only fits, in Ref.~\cite{Ball:2017otu} it has been studied the impact of the inclusion
of pseudo-data from possible future DIS experiments at LHeC and FCC-eh.
It is interesting to use those results to study the benefits that the construction of such experiments
may give in the prediction of the Higgs cross section.
While this study is interesting also beyond the business of small-$x$ resummation,
a striking feature of both the LHeC and FCC-eh datasets is to provide a significant reduction on the PDF uncertainty at small $x$,
also due to the extended sensitivity to smaller values of $x$ than reached at HERA,
thus also enhancing the sensitivity to small-$x$ resummation effects~\cite{Ball:2017otu}.
Most of the uncertainty reduction is provided by the FCC-eh dataset, with the LHeC dataset providing only an extra little improvement.
Being realistic (it is unlikely that both facilities will be built)
and also wanting to maximize the effect of the new data, we decide to consider the PDFs obtained
with the addition of the FCC-eh dataset alone.
In Fig.~\ref{fig:ggHkfactorDIS} (right plot) we show the relative PDF uncertainty of the N$^3$LO+LL result
for the real DIS-only fit (dashed blue) and the futuristic DIS-only fit including FCC-eh pseudo-data (dot-dashed green).
We see that indeed the reduction is significant and important in the high-energy region.
However, it is way less dramatic than the analogous reduction visible in the gluon PDF (see Ref.~\cite{Ball:2017otu}).
This is due to the fact that we are considering an inclusive cross section,
which, according to Eq.~\eqref{eq:sigmaCollzspace}, contains contributions from all the regions of $x$ from $\tau$ to $1$.
Thus, the strong uncertainty reduction on the gluon at small $x$ has only a limited benefit on the full PDF uncertainty
of the cross section even at rather large collider energies.
Indeed, for comparison, in the plot the uncertainty obtained with the global dataset (and thus without FCC-eh)
is also shown (solid red): this uncertainty is always smaller than the DIS-only with FCC-eh one,
up to the FCC-hh energy where they become comparable.
Thus, for the inclusive cross section, future DIS experiments may lead to an increased precision at high energies,
but also precise hadron collider data can, and only combining both of them one can achieve a higher precision.
When considering differential observables, which are more directly sensitive to the PDFs at specific values of the
momentum fraction, the uncertainty reduction provided by FCC-eh or LHeC may be more substantial.

\begin{figure}[t]
  \centering
  \includegraphics[width=0.243\textwidth,page=2]{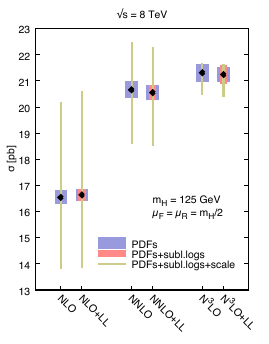}
  \includegraphics[width=0.243\textwidth,page=3]{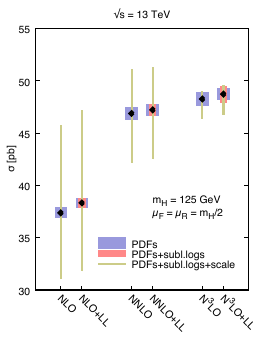}
  \includegraphics[width=0.243\textwidth,page=5]{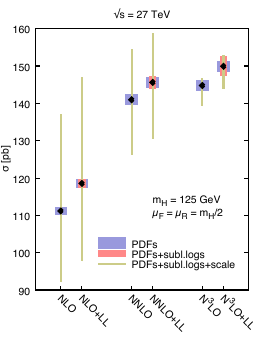}
  \includegraphics[width=0.243\textwidth,page=6]{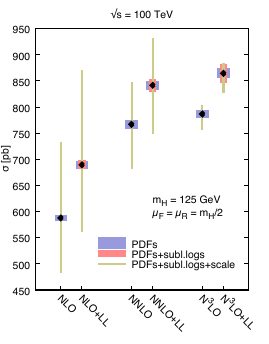}
  \caption{Perturbative progression of the Higgs cross section for four collider energies $\sqrt{s}=\left\{8,13,27,100\right\}$~TeV.
    In each plot the NLO, NLO+LL, NNLO, NNLO+LL, N$^3$LO and N$^3$LO+LL results are shown.
    The results are supplemented by uncertainty bands from PDF, subleading logarithms and scale uncertainties.}
  \label{fig:ggHpert}
\end{figure}
So far we have presented results (with and without resummation) at N$^3$LO.
To complete the discussion, we present some representative results at previous orders.
In Fig.~\ref{fig:ggHpert} we show the NLO, NNLO and N$^3$LO cross sections, and their counterparts with resummation,
for four choices of the collider energies, namely
$\sqrt{s}=8$~TeV (LHC Run 1),
$\sqrt{s}=13$~TeV (LHC Run 2),
$\sqrt{s}=27$~TeV (HE-LHC), and
$\sqrt{s}=100$~TeV (FCC-hh).
We use the global NNLO PDF set for all fixed-order predictions, and the global NNLO+NLL PDFs for all resummed predictions.
For each prediction we show various uncertainties.
At fixed-order, the PDF uncertainty (blue) and its sum in quadrature with the (asymmetric) scale uncertainty
(envelope of the standard 7-point scale variation, yellow).
At resummed level, the PDF uncertainty (blue), its sum in quadrature with the uncertainty from subleading logarithms (salmon),
and their sum in quadrature with the scale uncertainty (yellow).
We observe that since most of the effect of the resummation is due to the PDFs,
the increase in the cross section is more or less independent of the perturbative order.
Therefore the perturbative progression does not improve significantly when adding resummation,\footnote
{Threshold resummation, instead, has exactly the effect of predicting most of the higher order contributions,
and thus speeds up the perturbative convergence, see Refs.~\cite{Bonvini:2016frm,Bonvini:2018ixe}.}
even though a marginal improvement is anyway visible --- for instance,
at the FCC-hh the NLO full uncertainty band does not cover the NNLO result,
while the NLO+LL band does cover the central NNLO+LL result.
The scale uncertainty, being it dominated by the $\mur$ dependence, is not improved either,
again because most of the resummation effect is given by the PDFs, which only depend on $\muf$.
It is interesting to note that the uncertainty from subleading logarithmic contributions
is negligible at NLO+LL, small at NNLO+LL and quite large (comparable with scale uncertainty for HE-LHC and FCC-hh) at N$^3$LO+LL.
Because we compute this uncertainty as the difference between using NLL and LL$^\prime$ anomalous dimensions
in the resummation of coefficient functions, this pattern shows that these two approaches give quantitatively similar
results at NLO+LL and NNLO+LL, but as we have already noted they differ significantly at N$^3$LO+LL,
in agreement with the parton level results presented in Sect.~\ref{sec:ggHpartonic}.

We do not report explicit numerical results, as these have been already presented in Ref.~\cite{Bonvini:2018ixe},
where the contribution from threshold resummation is also included, which is known to stabilize the perturbative
expansion of the Higgs cross section, and additional corrections due to e.g.\ the bottom and charm quark
running in the loop are considered.
Therefore, the results of Ref.~\cite{Bonvini:2018ixe} are more appropriate for phenomenological applications.

We conclude the section with a final observation on the importance of considering the effect of small-$x$ resummation
for precision phenomenology.
At the current LHC energy including resummation leads to a $1\%$ increase of the cross section.
This effect is covered by the estimate of the theory uncertainty ${}^{+1.4\%}_{-3.6\%}$
from missing higher orders (in both coefficient functions and PDFs)
recommended by the LHC HXSWG~\cite{deFlorian:2016spz,Anastasiou:2016cez}.\footnote
{This uncertainty is the linear sum (as prescribed by Refs.~\cite{deFlorian:2016spz,Anastasiou:2016cez})
of the uncertainty from scale variations (${}^{+0.2\%}_{-2.4\%}$) and the estimate
of the uncertainty from missing higher order corrections in the PDFs ($\pm1.2\%$).}
However, when including additional corrections at threshold~\cite{Bonvini:2018ixe},
the overall effect of resummations becomes an increase of $2\%$ of the cross section,
which is outside the LHC HXSWG uncertainty.
This shows on the one hand that such uncertainty is likely underestimated,
and on the other hand that the inclusion of resummation(s) is \new{necessary } to achieve the (few) percent accuracy goal.
Moving to higher collider energies, \new{the effect of small-$x$ resummation becomes more substantial. }
For instance, we have seen that at the FCC-hh the effect of resummation amounts to
an increase of the cross section of approximately $10\%$.
\new{This is well outside the analogous estimate of the theory uncertainty from missing higher orders ${}^{+3.5\%}_{-4.6\%}$
presented in Ref.~\cite{Contino:2016spe},
mostly due to the fact that this estimate~\cite{Anastasiou:2016cez} of the uncertainty from missing higher orders in the PDFs
is only based on the perturbative progression at lower orders, and thus it
does not take into account the presence of logarithmically enhanced contributions at small $x$, which are responsible
for the sizeable effect of small-$x$ resummation to this cross section. }
Therefore, the inclusion of small-$x$ resummation is essential not only to reach a higher precision,
but also to avoid underestimating the potential effects of higher order corrections.
\new{These considerations easily hold for other processes as well, and in some cases
(e.g.\ differential observables more directly sensitive to small-$x$ PDFs)
these effects may be much more relevant even at the LHC.}

\section{Conclusions}
\label{sec:conclusions}

In this work we have extended the resummation formalism for partonic coefficient functions
originally developed for deep inelastic scattering~\cite{Bonvini:2016wki}
to the case of two hadron in the initial state, relevant for LHC.
In particular, at the leading logarithmic accuracy we considered,
only processes which are initiated by two gluons at LO,
such as Higgs production in gluon fusion, top-pair production, jet production, etc.,
are non-trivial, while processes which are quark initiated like Drell-Yan
resum only through a single initial state leg at this order,
and are thus treated identically to the single-hadron case.
We have demonstrated the equivalence of our (more general) approach with the original ABF approach
of Ref.~\cite{Ball:2007ra} under specific assumptions, and provided all the ingredients
needed to match resummed results to fixed-order computations up to N$^3$LO.
This formalism has been implemented in the new version of the public code \texttt{HELL 3.0}.

We then studied a specific hadron-hadron collider process, namely Higgs production in gluon fusion.
The partonic coefficient functions with incoming off-shell gluons needed for obtaining
the resummed on-shell coefficient functions for this process have been computed
a while ago~\cite{Marzani:2008az, DelDuca:2001fn,Pasechnik:2006du}.
However, it was possible to obtain consistent resummed predictions only thanks to two recent developments.
On the one hand, the creation of the public code \texttt{HELL} which implements the
formalism for resummation developed in Refs.~\cite{Bonvini:2016wki,Bonvini:2017ogt}
and extended to the hadron-hadron collider case in this work.
On the other hand, the existence of PDF sets which have been obtained using small-$x$ resummation
(from \texttt{HELL}) in their determination and evolution~\cite{Ball:2017otu,Abdolmaleki:2018jln}.

Comparing the Higgs cross section at N$^3$LO supplemented by small-$x$ resummation using
resummed NNLO+NLL PDFs with the (current standard according to the LHC HXSWG) fixed-order N$^3$LO
prediction using NNLO PDFs, we have found that the cross section increases mildly ($+1\%$) at current LHC energy,
and increases more substantially for larger collider energies,
reaching $+4\%$ at HE-LHC ($\sqrt{s}=27$~TeV)
and $+10\%$ at FCC-hh ($\sqrt{s}=100$~TeV).
In the $Q_0\MSbar$ scheme that we adopt, almost all of this effect comes from the use of resummed PDFs,
and in particular it is due to the fact that NNLO PDFs are unstable at small-$x$
due to the presence in the three-loop splitting functions of large unresummed logarithms of $x$~\cite{Ball:2017otu}.
The effect would be potentially much larger if (yet unavailable) N$^3$LO PDFs were used,
since four-loop splitting functions are even more unstable due to larger powers
of the logarithms at small-$x$~\cite{Bonvini:2018xvt}.

The main conclusion that we draw is that predictions based on NNLO PDFs and in future on N$^3$LO PDFs
will be unreliable for processes which are sensitive to small-$x$ PDFs,
due to the bias induced by the perturbative instability of the splitting functions
\new{and coefficient functions of processes used for PDF determination,
which is not accounted for in the way PDF uncertainties are estimated. }
While for the inclusive Higgs cross section this seems to be the case only at future colliders,
for differential observables which are more directly sensitive to PDFs at a given momentum fraction
this conclusion \new{may } hold also at the LHC in specific kinematic configurations (e.g., large rapidities).
In these cases, \new{a reassessment of the PDF uncertainties at small-$x$ is mandatory,
for instance by comparing theoretical predictions obtained with PDF sets with and without small-$x$ resummation.
The most reliable theoretical predictions should, in these cases, be based on small-$x$ resummed computation.}

At the moment, the main limitation of small-$x$ resummation is its limited logarithmic accuracy.
For DGLAP evolution, resummation is known at NLL, while for the coefficient functions
it is known only at LL.
In this work we have also studied the potential effect of subleading logarithmic contributions
to the Higgs cross section, by computing different theoretical predictions which differ by subleading
terms both in the coefficient functions and in the PDFs.
The effect is potentially large, and while the qualitative conclusions of this study remain unchanged,
achieving high precision requires the extension of the small-$x$ resummation formalism to higher logarithmic order.
This ambitious goal is left to future work.

The new \texttt{3.0} version of \texttt{HELL} which contains all these new developments
is publicly available for download at the address
\begin{center}
  \href{http://www.ge.infn.it/~bonvini/hell}{\tt www.ge.infn.it/$\sim$bonvini/hell}
\end{center}
It also uses a new default for the implementation of the resummation,
as discusses in Ref.~\cite{Bonvini:2018xvt}.
\texttt{HELL 3.0} has been used in Ref.~\cite{Bonvini:2018ixe} to obtain double-resummed predictions
at threshold (large $x$) and at high energy (small $x$) for the Higgs cross section at LHC and beyond.

\acknowledgments
{
I'm grateful to Simone Marzani for his encouragement to write this paper and for uncountable discussions,
since part of this work was instrumental for Ref.~\cite{Bonvini:2018ixe}.
I also want to thank Richard Ball, Stefano Forte, Giovanni Ridolfi, Juan Rojo and Luca Rottoli for various discussions
and for their feedback on the manuscript.
This work is supported by the Marie Sk\l{}odowska Curie grant HiPPiE@LHC.
}

\appendix
\section{Off-shell coefficient function for Higgs production}
\label{sec:offshell}

In this appendix we report some expressions which are needed for the actual computation
of the resummed coefficient functions for $ggH$. In particular, we report the off-shell coefficient function,
we explain how one can conveniently change variables for obtaining a reliable numerical integration,
and we show how the $M$-expansion coefficient of the Mellin transform of the off-shell coefficient function
(needed for the perturbative expansion of the resummed results) can be constructed.
We stress that all the details given in this appendix, with the exception of the explicit
expressions of the off-shell coefficients, are very general and can be used for other processes
with two incoming off-shell gluons as well.

\subsection{Off-shell coefficient function in suitable variables}
\label{sec:IFchangeofvariables}

The lowest order off-shell coefficient function for $ggH$ production
with both gluons off-shell has been computed in Ref.~\cite{Marzani:2008az}.
Its form is
\beq\label{eq:Coff_f12}
{\cal C}\(0, \xi_1, \xi_2,\as\) = f_1(\xi_1,\xi_2) + \xi_1\xi_2 f_2(\xi_1,\xi_2) +\Ord(\as),
\eeq
with
\begin{align}
f_1(\xi_1,\xi_2) &= \frac{\abs{A_1}^2}{4\abs{1-\frac14(1-4/\rhot)s_0^2(\rhot)}^2} ,\\
f_2(\xi_1,\xi_2) &= \frac{2\abs{A_3}^2}{4\abs{1-\frac14(1-4/\rhot)s_0^2(\rhot)}^2} ,\\
s_0(\rhot) &=
\begin{cases}
\log\frac{1-\sqrt{1-4/\rhot}}{1+\sqrt{1-4/\rhot}} +i\pi & \rhot>4\\
2i\sin^{-1}\sqrt{\rhot/4} & \rhot\geq4,
\end{cases}
\end{align}
and we recall that $\rhot=\mh^2/\mt^2$.
The dimensionless form factors $A_1$ and $A_3$ have been computed in Refs.~\cite{DelDuca:2001fn,Pasechnik:2006du}.
Before presenting their form, we observe that numerical integration of this function
is problematic in the region $\xi_1\sim\xi_2$. Since the off-shell cross section is symmetric
under the exchange of virtualities $\xi_1\leftrightarrow\xi_2$, we suggest the change of variables
\beq\label{eq:changevariables}
\xi_1 = t(1+y),\qquad \xi_2= t(1-y).
\eeq
Thus, the integral over virtualities of a function $F(\xi_1,\xi_2)$ transforms as
\begin{align}
\int_0^\infty d\xi_1 \int_0^\infty d\xi_2\, F(\xi_1,\xi_2)
&= \int_{-1}^1dy \int_0^\infty dt \, 2t \, F(t(1+y), t(1-y)) \nonumber\\
&= 2\int_0^1dy \int_0^\infty dt \, 2t \, F(t(1+y), t(1-y)),
\end{align}
where in the last line we have assumed $F$ to be symmetric, so that the problematic
region $\xi_1=\xi_2$ lies at the boundary of the integration domain and can be better integrated numerically.
In terms of these variables, the form factors~\cite{DelDuca:2001fn,Pasechnik:2006du} have a simpler form given by
\begin{align}
  A_1(t,y)
  &= \frac{C_0(t,y)}{\Delta_3}\[\frac{4(1+2t)}{\rhot} -(1+2t)^2 + 12\frac{(1+2t)t^2(1-y^2)}{\Delta_3} \]
  + 2\frac{1+2t}{\Delta_3} \nonumber\\
  &\quad + \frac2{\Delta_3} \Big[B_0\big(-t (1+y)\big) + B_0\big(-t (1-y)\big) - 2B_0(1)\Big]
  \[ t - 6\frac{t^2(1-y^2)}{\Delta_3} \] \nonumber\\
  &\quad + \frac{2ty}{\Delta_3} \Big[B_0\big(-t (1+y)\big) - B_0\big(-t (1-y)\big)\Big] \[ 1 + 12\frac{t^2(1-y^2)}{\Delta_3} \],
  \label{eq:A1}\\
  A_3(t,y)
  &= \frac{C_0(t,y)}{\Delta_3} \[\frac{8}{\rhot} -4-4t + 6\frac{(1+2t)^2}{\Delta_3} \] + \frac4{\Delta_3} \nonumber\\
  &\quad + \Big[B_0\big(-t (1+y)\big) + B_0\big(-t (1-y)\big) - 2B_0(1)\Big] \frac{2}{\Delta_3}\(1-3\frac{1+2t}{\Delta_3}\) \nonumber\\
  &\quad + \Big[B_0\big(-t (1+y)\big) - B_0\big(-t (1-y)\big)\Big] \frac{12ty(1+2t)}{\Delta_3^2},\label{eq:A3}
\end{align}
 with\footnote{We write $C_0$ in the form given in Ref.~\cite{Pasechnik:2006du}, which turns out to be numerically much more stable.}
\begin{align}
  \Delta_3 &= 1+4t + 4t^2 y^2\\
  B_0(\zeta) &= -\sqrt{\frac{\zeta-4/\rhot}{\zeta}}\log\frac{\sqrt{\frac{\zeta-4/\rhot}{\zeta}}+1}{\sqrt{\frac{\zeta-4/\rhot}{\zeta}}-1}, \\
  C_0(t,y) &= \frac1{\sqrt{\Delta_3}} \Big[ \kappa(\delta_0,T_0) + \kappa(\delta_+,T_+) + \kappa(\delta_-,T_-) \Big],\label{eq:C0}\\
  \kappa(\delta,T) &= \Li_2\(\frac{\delta-1}{\delta-T}\) + \Li_2\(\frac{\delta-1}{\delta+T}\)
  - \Li_2\(\frac{\delta+1}{\delta-T}\) - \Li_2\(\frac{\delta+1}{\delta+T}\)
\end{align}
and
\begin{align}
  \delta_0 &= \frac{1+2t}{\sqrt{\Delta_3}} &
  T_0 &= \sqrt{1-4/\rhot}\\
  \delta_\pm &= -\frac{1\pm 2ty}{\sqrt{\Delta_3}} &
  T_\pm &= \sqrt{1+\frac{4/\rhot}{t(1\pm y)}}.
\end{align}
All these expressions have been coded in \texttt{HELL 3.0}.
In some particular limits, where some of the functions fail to evaluate numerically
(mostly due to the square root terms),
Taylor expansions are used to overcome this problem.

In the actual definition of the resummed coefficient functions, Eqs.~\eqref{eq:Cggaux1} and \eqref{eq:Cggaux2},
the integration extends from the position of the Landau pole $\xi_0$ to infinity,
and the off-shell coefficient appears with derivatives with respect to $\xi_1$ and $\xi_2$.
The second fact is per se not a problem, except that these derivatives must be computed analytically
both for speed reasons and to avoid proliferation of numerical errors.
Therefore, it is useful to limit as much as possible the number of derivatives to be computed.
To do so, we first observe that we do not need to treat identically the contributions
from $f_1$ and $f_2$, Eq.~\eqref{eq:Coff_f12}.
Indeed, in our numerical implementation we use the expression in which the derivatives act on the coefficient
function for the $f_1$ contribution, while we use the one with derivatives on the evolution functions for the $f_2$ contribution.
Making the notation very schematic and omitting all arguments except the virtualities,
we write Eq.~\eqref{eq:Cggaux1} as
\beq\label{eq:Cgg_}
C_{gg}
= \int d\xi_1 \int d\xi_2\, 
\[U(\xi_1) U(\xi_2) \frac{\partial^2 f_1 \(\xi_1,\xi_2\)}{\partial \xi_1\partial \xi_2}
+ U'(\xi_1) U'(\xi_2)\xi_1\xi_2 f_2\(\xi_1,\xi_2\)\],
\eeq
where $U$ is a shorthand for $U_{\rm ABF}^{\rm ht}$, and $U'(\xi)$ is its the derivative with respect to $\xi$.
The term proportional to $f_2$ is then treated as described above, namely by changing variables according
to Eq.~\eqref{eq:changevariables} and using the symmetry to integrate only for positive $y$'s.
The contribution to $C_{gg}$ from $f_1$, which we call $C_1$ for simplicity,
is instead manipulated as follows
\begin{align}
C_1
  &\equiv \int d\xi_1 \int d\xi_2\, U(\xi_1) U(\xi_2) \frac{\partial^2 f_1 \(\xi_1,\xi_2\)}{\partial \xi_1\partial \xi_2}\nonumber\\
  &= \frac12 \int dy \int dt \, U_+\,U_- 
  \bigg[
    t\frac{\partial^2 f_1}{\partial t^2}
    -2y\frac{\partial^2f_1}{\partial t\partial y}
    -\frac1t \frac{\partial}{\partial y}\((1-y^2)\frac{\partial f_1}{\partial y}\)
  \bigg]
  \nonumber\\
  &= -\frac12 \int dy \int dt \, U_+\,U_- 
  \bigg[
    2y\frac{\partial^2f_1}{\partial t\partial y}
    +\(1+t(1+y)\frac{U_+'}{U_+}+t(1-y)\frac{U_-'}{U_-}\)\frac{\partial f_1}{\partial t}
\nonumber\\ &\hspace{8cm}
    -\(\frac{U_+'}{U_+}-\frac{U_-'}{U_-}\)(1-y^2)\frac{\partial f_1}{\partial y}
  \bigg]
\label{eq:C1}
\end{align}
where we have defined
\beq
U_\pm \equiv U(t(1\pm y)),
\eeq
and $U_\pm'$ are still derivatives with respect to the full argument.
In the first step in Eq.~\eqref{eq:C1} we have simply performed the change of variables;
in the second step we have integrated by parts some contributions to remove double $t$ and double $y$ derivatives
(all boundary terms vanish).
At this point one can use the symmetry $y\to-y$ to restrict the integration to positive $y$'s up to an overall factor of 2.
Eq.~\eqref{eq:C1} is what we use in the code, and provides a stable numerical evaluation of the integral,
with the advantage of depending on a single second derivative of the off-shell coefficient function.

The auxiliary function Eq.~\eqref{eq:Cggaux2} is instead much simpler to treat.
First, the $f_2$ term proportional to $\abs{A_3}^2$ does not contribute, since it is multiplied by $\xi_1\xi_2$
and one of them is zero (say, $\xi_2=0$), so we have
\beq\label{eq:Caux_}
C_{\rm aux}
= \int d\xi\, U(\xi) \frac{\partial f_1 \(\xi,0\)}{\partial \xi}.
\eeq
Second, there is a single derivative, which can be directly obtained from $\partial f_1/\partial t$ used above.
In fact, the form factor becomes much simpler in the limit $\xi_2=0$, i.e.\ $y=1$,
\beq
A_1(t,1)
= C_0(t,1)\[\frac{4/\rhot}{1+2t} -1\]
+ \frac2{1+2t}
+ \frac{4t}{(1+2t)^2} \Big[B_0(-2t) - B_0(1)\Big]
\eeq
with
\beq
C_0(t,1) = \frac{1}{1+2t}\[
  \Li_2\(\frac2{1+T_+}\) + \Li_2\(\frac2{1-T_+}\)
- \Li_2\(\frac2{1+T_0}\) - \Li_2\(\frac2{1-T_0}\)
\],
\eeq
being now $T_+ = \sqrt{1+2/(\rhot t)}$. This analytical expression is also useful for cross-checking numerically
part of the results used above in the $C_{gg}$ case.

We can now discuss the implication of restricting the integration to $\xi_{1,2}>\xi_0=\exp\frac{-1}{\as\beta_0}$.
Let us start with the one-dimensional case, $C_{\rm aux}$, Eq.~\eqref{eq:Caux_}.
The integrand is peaked at $\xi\sim\muf^2/Q^2\sim1$, and drops at large $\xi$ as a negative power of $\xi$
(in this case, as $1/\xi^3$).
Thus, we do not loose precision if we approximate the integrand as
\beq
\int_{\xi_0}^\infty d\xi\, F(\xi) \simeq \int_{\xi_0}^{\xi_F^{m+1}\xi_0^{-m}} d\xi\, F(\xi),
\qquad \xi_F = \frac{\muf^2}{Q^2},
\eeq
where $m>0$ cuts off the large $\xi$ region which gives a negligible contribution to the integral,
and $F(\xi)$ is a generic name for the integrand.
In practice, we have noticed that $m=3$ is sufficiently small to guarantee numerical stability and at the same time
sufficiently large to keep the important region of the integral and cut away only negligible corrections.
We then split the integrand in two pieces, from $\xi_0$ to $\xi_F$ and from $\xi_F$ to $\xi_F^{m+1}\xi_0^{-m}$,
and perform the change of variables $\xi=\xi_F\exp(-u)$ and $\xi=\xi_F\exp(mu)$ respectively:
\beq
\int_{\xi_0}^{\xi_F^{m+1}\xi_0^{-m}} d\xi\, F(\xi) = \int_0^{\log(\xi_F/\xi_0)} du\, \xi_F \[e^{-u}F(\xi_Fe^{-u}) + m e^{mu}F(\xi_Fe^{mu})\].
\eeq
Then we can change variable again according to $u=v\log(\xi_F/\xi_0) = \frac{v}{\beta_0\as(\muf^2)}$, and get finally
\beq
\int_{\xi_0}^\infty d\xi\, F(\xi) \simeq
\int_0^1 \frac{dv}{\as\beta_0} \,\xi_F\[e^{\frac{-v}{\as\beta_0}} F\(\xi_F e^{\frac{-v}{\as\beta_0}}\)
  + m e^{\frac{mv}{\as\beta_0}} F\(\xi_F e^{\frac{mv}{\as\beta_0}}\)\],
\eeq
such that the integration region is in the unit hypercube (of dimension 1 in this case),
and thus directly usable in standard numerical integration routines.
This expression is what is used in \texttt{HELL} for the one-dimensional case.

In the two-dimensional case, we need to convert the two conditions
$\xi_1>\xi_0$ and $\xi_2>\xi_0$ into a condition for the $y,t$ integration range.
Assuming to integrate in $y$ first, the condition becomes
\beq
-\(1-\frac{\xi_0}t\) < y < 1-\frac{\xi_0}t,\qquad t>\xi_0.
\eeq
The integral of a generic function $F(t,y)$, once the $y$ integration is symmetrized,
can be treated as in the one-dimensional case, approximating the integral and performing subsequent changes of variables,
\begin{align}\label{eq:HiggsHELLintegral}
\int_{\xi_0}^\infty dt \int_0^{1-\xi_0/t} dy \, F(t,y)
&\simeq \int_{\xi_0}^{\xi_F^{m+1}\xi_0^{-m}}dt \int_0^{1-\xi_0/t} dy \, F(t,y)
\\
&= \int_0^{\log(\xi_F/\xi_0)} du\,\xi_F \bigg[e^{-u}\int_0^{1-\frac{\xi_0}{\xi_F}e^{u}} dy \, F(\xi_F e^{-u},y) 
\nonumber\\
  &\qquad\qquad\qquad\qquad\quad+ m e^{mu}\int_0^{1-\frac{\xi_0}{\xi_F}e^{-mu}} dy \, F(\xi_F e^{mu},y) \bigg]
\nonumber\\
&= \int_0^1 \frac{dv}{\as\beta_0} \int_0^1 dw\,\xi_F \bigg[e^{\frac{-v}{\as\beta_0}}\(1-e^{\frac{v-1}{\as\beta_0}}\) F\(\xi_F e^{\frac{-v}{\as\beta_0}}, \(1-e^{\frac{v-1}{\as\beta_0}}\) w\)
\nonumber\\
  &\qquad\qquad\qquad\qquad+ m e^{\frac{mv}{\as\beta_0}}\(1-e^{\frac{-mv-1}{\as\beta_0}}\) F\(\xi_F e^{\frac{mv}{\as\beta_0}}, \(1-e^{\frac{-mv-1}{\as\beta_0}}\)w\) \bigg],\nonumber
\end{align}
where in the last step we first changed variable
$y=(1- e^{u}  \xi_0/\xi_F)w$ in the first $y$ integral and
$y=(1- e^{-mu}\xi_0/\xi_F)w$ in the second $y$ integral,
and then we used again $u=v\log(\xi_F/\xi_0)$.
As before, the final result is integrated in the unit hypercube (of dimension 2 in this case),
and thus immediately usable for numerical integration.
This expression is implemented in \texttt{HELL} for the two-dimensional case.

\subsection{Impact factor and its expansion coefficients}
\label{sec:IFcoeff}

We now move to the computation of the coefficients of the $M_{1,2}$ expansion
of the Mellin transform of the off-shell coefficient function, Eq.~\eqref{eq:impactfactor}.
Such Mellin transform is equivalent to Eq.~\eqref{eq:Cgg_} after replacing
\beq\label{eq:replacement}
U(\xi_i) \to \(\frac{Q^2}{\muf^2}\xi_i\)^{M_i},\qquad i=1,2
\eeq
and letting $\xi_0\to0$.
We can thus start from Eq.~\eqref{eq:C1} and, after integrating by parts in $t$ the last term, we arrive at
(again omitting all non-crucial arguments)\footnote{Note that the integrand is no longer symmetric for 
$y\to-y$, unless $M_1$ and $M_2$ (which keep reference to the incoming gluon legs) are swapped simultaneously.}
\begin{align}\label{eq:CoffMell1}
  \tilde{\cal C}(M_1,M_2)
  &= -\frac12 \int_{-1}^1 dy \int_0^\infty dt\,\(\frac{Q^2}{\muf^2}t\)^{M_1+M_2}(1+y)^{M_1}(1-y)^{M_2}
  \nonumber\\ &\qquad \times
  \[
  y\frac{\partial^2 f_1}{\partial t \partial y}
  +(1+M_1+M_2)\frac{\partial f_1}{\partial t}
  +\frac{M_1-M_2}{M_1+M_2} \frac{\partial^2f_1}{\partial t\partial y}
  -4M_1M_2 tf_2
  \].
\end{align}
Our goal is now to expand this expression in powers of $M_1$ and $M_2$, to construct the coefficients $\tilde{\cal C}_{kj}$,
Eq.~\eqref{eq:impactfactorser}.
We observe however that there is a term in Eq.~\eqref{eq:CoffMell1} which seems to give rise to negative powers
of $M_{1,2}$, namely the one with $M_1+M_2$ in the denominator.
When expanding $(tQ^2/\muf^2)^{M_1+M_2}$ in powers of $M_1+M_2$ all terms except the zero-th order term
will compensate the denominator. Thus, the only term which is potentially dangerous is the zero-th order one, which reads
\beq
-\frac12 \int_{-1}^1 dy \,(1+y)^{M_1}(1-y)^{M_2}
\frac{M_1-M_2}{M_1+M_2} \int_0^\infty dt\,\frac{\partial^2f_1}{\partial t\partial y}.
\eeq
But this term vanishes, since
\beq
\int_0^\infty dt\,\frac{\partial^2f_1}{\partial t\partial y}
= -\left.\frac{\partial f_1}{\partial y}\right|_{t=0} = 0,
\eeq
because $f_1$ in $t=0$ is independent of $y$.
This proves that only non-negative powers of $M_{1,2}$ are produced in the expansion of Eq.~\eqref{eq:CoffMell1}, as it must.
To compute the coefficients of such an expansion in a systematic way, we find it convenient to introduce the variables
\beq
M_\pm = \frac{M_1\pm M_2}2,
\eeq
in terms of which Eq.~\eqref{eq:CoffMell1} becomes
\begin{align}\label{eq:CoffMell2}
  \tilde{\cal C}(M_1,M_2)
  &= -\frac12 \int_{-1}^1 dy \int_0^\infty dt\, \(\frac{Q^2}{\muf^2}t\)^{2M_+}(1-y^2)^{M_+} \(\frac{1+y}{1-y}\)^{M_-}
  \nonumber\\ &\qquad \times
  \[
  \frac{\partial}{\partial y}\(y\frac{\partial f_1}{\partial t}\)
  +2M_+\frac{\partial f_1}{\partial t}
  +\frac{M_-}{M_+} \frac{\partial^2f_1}{\partial t\partial y}
  -4(M_+^2-M_-^2)t f_2
  \] \nonumber\\
  &= -\frac12 \int_0^1 dy \int_0^\infty dt\, \(\frac{Q^2}{\muf^2}t\)^{2M_+}(1-y^2)^{M_+} 
  \nonumber\\ &\qquad \times \Bigg\{
  \[\(\frac{1+y}{1-y}\)^{M_-}+\(\frac{1-y}{1+y}\)^{M_-}\]\[
  \frac{\partial}{\partial y}\(y\frac{\partial f_1}{\partial t}\)
  +2M_+\frac{\partial f_1}{\partial t}
  -4(M_+^2-M_-^2)t f_2
  \]
  \nonumber\\ &\qquad\quad
  +\[\(\frac{1+y}{1-y}\)^{M_-}-\(\frac{1-y}{1+y}\)^{M_-}\]
  \frac{M_-}{M_+} \frac{\partial^2f_1}{\partial t\partial y}
  \Bigg\},
\end{align}
where in the second step we have symmetrized the integration in $y$ and restricted it to positive $y$'s.
Defining 
\beq
L_+ = \log\[t^2(1-y^2)\] + 2\log\frac{Q^2}{\muf^2},\qquad
L_- = \log\(\frac{1+y}{1-y}\),
\eeq
we can expand Eq.~\eqref{eq:CoffMell2} as
\begin{align}
  \tilde{\cal C}(M_1,M_2)
  &= - \int_0^1 dy \int_0^\infty dt\,
  \sum_{a=0}^\infty \frac{M_+^{a}}{a!} L_+^{a}
  \sum_{b=0}^\infty \frac{M_-^{b}}{b!} L_-^{b}
    \,\frac{1+(-1)^{b}}2
  \nonumber\\ &\qquad \times \Bigg\{
  \frac{\partial}{\partial y}\(y\frac{\partial f_1}{\partial t}\)
  +2M_+\frac{\partial f_1}{\partial t}
  + \frac{b}{M_+L_-} \frac{\partial^2f_1}{\partial t\partial y}
  -4(M_+^2-M_-^2)t f_2
  \Bigg\} \nonumber\\
  &= \sum_{a=0}^\infty \sum_{\substack{b=0\\b\,\rm even}}^\infty M_+^{a} M_-^{b} c_{a,b},
\end{align}
where the coefficients $c_{a,b}$ of the $M_\pm$ expansion are given by
\begin{align}\label{eq:coeff+-}
  c_{a,b}
  = \frac{1}{a!b!} \int_0^1 dy \int_0^\infty dt\,\Bigg\{
  & - L_+^{a} L_-^{b} \frac{\partial}{\partial y}\(y\frac{\partial f_1}{\partial t}\)
  - 2 a L_+^{a-1} L_-^{b} \frac{\partial f_1}{\partial t}
  - \frac{b}{a+1} L_+^{a+1} L_-^{b-1} \frac{\partial^2f_1}{\partial t\partial y}
  \nonumber\\ &
  + 4a(a-1) L_+^{a-2} L_-^{b} t f_2
  - 4b(b-1) L_+^{a} L_-^{b-2} t f_2
  \Bigg\}.
\end{align}
Once these coefficients are known, they can be converted to the desired coefficients
$\tilde{\cal C}_{kj}$, Eq.~\eqref{eq:impactfactorser}, through the relation
\begin{align}
\tilde{\cal C}_{k,j}
&= \frac1{2^{k+j}} \sum_{\substack{b=0\\b\,\rm even}}^{k+j} c_{k+j-b,b}
\sum_{i=\max(0,b-j)}^{\min(b,k)} (-1)^i \binom{b}{i} \binom{k+j-b}{k-i}.
\end{align}
The integrals defining the coefficients Eq.~\eqref{eq:coeff+-} are suitable for numerical evaluation.
We stress that a straightforward expansion in powers of $M_{1,2}$ of Eq.~\eqref{eq:Cgg_}
after the replacement Eq.~\eqref{eq:replacement} suffers from a definition of the coefficients $\tilde{\cal C}_{kj}$
in terms of integrals that are not easy to perform numerically and give rise to large numerical errors.
Therefore, our construction, despite being somewhat involved, has the big advantage of reducing
the numerical error significantly, which was possible by exploiting the symmetry of the off-shell coefficient function.
We add that the construction presented in this subsection was actually already used for computing these coefficients
for Ref.~\cite{Ball:2013bra}, but it is presented in this detail here for the first time.

\phantomsection
\addcontentsline{toc}{section}{References}
\bibliographystyle{jhep}
\bibliography{references}

\end{document}